\documentclass{elsarticle}
\pdfoutput=1
\usepackage{amsmath}
\usepackage{amsfonts} 
\usepackage[super]{nth} 
\usepackage[top=1in, bottom=1.in, left=1.in, right=1.in]{geometry}
\usepackage{caption}

\usepackage{tikz}
\usepackage{ifthen} 
\newcommand*{\xMin}{0}%
\newcommand*{\xMax}{13}%
\newcommand*{\yMin}{-4}%
\newcommand*{\yMax}{3}%

\usepackage{bbm} 
\usepackage{subcaption} 
\usepackage{booktabs} 
\usepackage{placeins} 
\usepackage{hyperref} 
\usepackage{mathtools} 

\usepackage{etoolbox}
\let\theparentequation\theequation
\patchcmd{\theparentequation}{equation}{parentequation}{}{}
\renewenvironment{subequations}[1][]{
  \refstepcounter{equation}%
  \setcounter{parentequation}{\value{equation}}
  \setcounter{equation}{0}
  \def\theequation{\theparentequation\alph{equation}}%
  \let\parentlabel\label
  \ifx\\#1\\\relax\else\label{#1}\fi
  \ignorespaces
}{%
  \setcounter{equation}{\value{parentequation}}
  \ignorespacesafterend
}
\newcommand*{\nextParentEquation}[1][]{
  \refstepcounter{parentequation}
  \setcounter{equation}{0}
  \ifx\\#1\\\relax\else\parentlabel{#1}\fi
}

\newcommand{\half}{\frac{1}{2}}
\renewcommand{\l}{\left}
\renewcommand{\r}{\right}

\newcommand{\refeqn}[1]{Eq.~\eqref{eqn:#1}}
\newcommand{\refeqns}[1]{Eqs.~\eqref{eqns:#1}}
\newcommand{\mcone}{\mathbbm{1}}

\newcommand{\tvp}{\tilde{\varphi}}
\newcommand{\eps}{\epsilon}

\newcommand{\bOmega}{\boldsymbol{\Omega}}
\newcommand{\bG}{\boldsymbol{G}}
\newcommand{\bq}{\boldsymbol{q}}

\newcommand{\bx}{\boldsymbol{x}}

\allowdisplaybreaks

\begin{document}
\title{A Quadratic Programming Flux Correction Method for \\ High-Order DG Discretizations of \texorpdfstring{$S_N$}{SN} Transport}

\author[add1]{Ben C. Yee}\corref{cor1}
\ead{yee26@llnl.gov}
\author[add1,add2]{Samuel S. Olivier}
\ead{solivier@berkeley.edu}
\author[add1]{Terry S. Haut}
\ead{haut3@llnl.gov}
\author[add1]{Milan Holec}
\ead{holec1@llnl.gov}
\author[add1]{Vladimir Z. Tomov}
\ead{tomov2@llnl.gov}
\author[add3]{Peter G. Maginot}
\ead{pmaginot@llnl.gov}

\address[add1]{Lawrence Livermore National Laboratory \\
  7000 East Avenue, Livermore, CA 94550}
\address[add2]{Applied Science and Technology, University of California, Berkeley \\
  Berkeley, CA 94708}
\address[add3]{Los Alamos National Laboratory \\
  P.O. Box 1663, Los Alamos, NM 87545 }

\cortext[cor1]{Corresponding author}

\begin{abstract}
	We present a new flux-fixup approach for arbitrarily high-order discontinuous Galerkin (DG) discretizations of the $S_N$ transport equation, and we demonstrate the compatibility of this approach with the Variable Eddington Factor (VEF) method~\cite{olivier2017variable,olivier2019high}.
The new fixup approach is sweep-compatible: during a transport sweep (block Gauss-Seidel iteration in which the scattering source is lagged), a local quadratic programming (QP) problem is solved in each spatial element to ensure that the solution satisfies certain physical constraints, including local particle balance.
In this paper, we describe two choices of physical constraints, resulting in two variants of the method: QP Zero (QPZ) and QP Maximum Principle (QPMP).
In QPZ, the finite element coefficients of the solution are constrained to be nonnegative.  In QPMP, they are constrained to adhere to an approximate discrete maximum principle.

There are two primary takeaways in this paper.
First, when the positive Bernstein basis is used for DG discretization, the QPMP method eliminates negativities, preserves high-order accuracy for smooth problems, and significantly dampens unphysical oscillations in the solution.
The latter feature -- the dampening of unphysical oscillations -- is an improvement upon standard, simpler fixup approaches such as the approach described in~\cite{hamilton2009negative} (denoted as the ``zero and rescale'' (ZR) method in this paper).
This improvement comes at a moderate computational cost, but it is not prohibitive.  Our results show that, even in an unrealistic worst-case scenario where 83\% of the spatial elements require a fixup, the computational cost of performing a transport sweep with fixup is only $\sim$31\% greater than performing one without fixup.

The second takeaway is that the VEF method can be used to accelerate the convergence of transport sweeps even when a fixup is applied.
When optically thick regions are present, transport sweeps converge slowly, regardless of whether a fixup is applied, and acceleration is needed.
However, attempting to apply standard diffusion synthetic acceleration (DSA) to fixed-up transport sweeps results in divergence for optically thick problems.
Our results show that the same is not true for VEF.
When VEF is combined with fixed-up transport sweeps, the result is a scheme that produces a nonnegative solution, converges independently of the mean free path, and, in the case of the QPMP fixup, adheres to an approximate discrete maximum principle.
	
\end{abstract}

\maketitle

\section{INTRODUCTION}

Simulations of astrophysical phenomena and inertial confinement fusion (ICF) require the simultaneous numerical modeling of hydrodynamics and thermal radiative transfer (TRT), the latter of which typically includes the discrete-ordinates ($S_N$) transport equation.
In hydrodynamics, high-order discretizations and curved meshes are useful for resolving the complex movements of the material.
Unlike low-order methods, high-order methods provide greater robustness (especially in the presence of significant mesh distortions), symmetry preservation, and excellent strong scaling~\cite{langer2015performance,dobrev2012high,anderson2018high}.
The multiphysics nature of ICF and astrophysics problems introduce significant coupling between the hydrodynamics and TRT equations, and the robustness and accuracy of a multiphysics model is typically restricted by its least robust or least accurate component.
Because of this, high-order methods and curved meshes are also desired for the transport equation.  Until recently~\cite{haut2019efficient,haut2018dsa}, there has not been significant research in the development of higher-order transport discretizations on curved meshes.  Aside from a few works such as \cite{wang2009high,wang2010diffusion,maginot2016high,woods2018discrete,woods2019rz}, previous work in transport has primarily focused on \nth{1}- or \nth{2}-order spatial discretizations such as method of characteristics, linear discontinuous, or diamond difference.
Although these low-order methods are sufficient for many applications, they are not ideal for our aforementioned applications of interest.

One significant challenge of using high-order methods is that they are prone to negativities in under-resolved regions of the solution, especially near material boundaries and other discontinuities.  For any physical problem (nonnegative sources and boundary conditions), the continuous $S_N$ transport equations yield nonnegative solutions.  Thus, an ideal numerical scheme for the transport equation should be \emph{positivity-preserving}.  However, it is well known that there is a fundamental tradeoff between accuracy and positivity~\cite{lathrop1969spatial}, and negativities are nearly unavoidable in high-order schemes because it is typically not computationally feasible to refine the mesh to fully resolve every region of the domain and align with every discontinuity.
Despite this drawback, high-order methods are still desired for transport.
Without high-order transport, one would have to interpolate fields between a high-order hydrodynamics mesh and a low-order transport mesh.
Attempts to ``straighten out'' high-order curved meshes with low-order mesh elements can significantly increase the number of degrees of freedom (DOFs) in the already memory intensive 7-dimensional $S_N$ transport equations.
Moreover, mapping between high-order and low-order meshes is likely to reduce robustness, stability, and physics fidelity~\cite{haut2019efficient}.

Thus, a negativity correction or ``fixup'' is needed.  Aside from aesthetic concerns, negative solutions are particularly problematic for multiphysics simulations, as any of the numerical physics packages may fail due to unphysical inputs from other physics packages.
In TRT problems, negative intensities (scalar fluxes) produce negative absorption terms in the material energy balance equation, significantly increasing the likelihood of negative temperatures.
When negative temperatures occur, evaluations of opacities are not well-defined and the Planckian integral diverges.  This causes convergence issues for many nonlinear iterative solvers.

Previous development of positivity-preserving schemes for multidimensional transport primarily falls in at least one of the two following categories: fixups for low-order schemes such as diamond difference~\cite{lewisbook}, and computationally expensive nonlinear/nonpolynomial spatial discretizations~\cite{wareing_exp,maginot_LD,maginot2017nonnegative}.
In other words, the development of positivity-preserving schemes for high-order finite element discretizations of transport has been limited.
In \cite{maginot2015lumping}, lumping techniques are used to mitigate the appearance of negativities, but they do not ensure strict nonnegativity.  In~\cite{yuan2016high}, high-order solutions are corrected using a combination of scaling and rotational positivity-preserving ``limiters'' to eliminate negativities.  However, the scaling limiter cannot be used when the element-average is negative, while the rotational limiter does not preserve particle balance (\nth{0} spatial moment of the transport equation).  These limiters can also be difficult to implement in higher dimensions.  \cite{hamilton2009negative} describes a simple, sweep-compatible flux-fixup.  In each spatial element, negativities are zeroed, and the remaining components are rescaled to preserve particle balance.   Though it is applied only to linear discontinuous discretizations in~\cite{hamilton2009negative}, it is straightforward to generalize the approach to arbitrarily high-order discontinuous Galerkin (DG) methods.  This ``zero and rescale'' (ZR) method serves as a point of reference for evaluating the performance of our new methods.

In this work, we present a new, quadratic programming (QP) based flux-fixup approach for arbitrarily high-order DG discretizations of $S_N$ transport.
Like the ZR method, our QP-based approach preserves particle balance and is sweep-compatible -- fixups are applied by solving local QP problems in each spatial element during the transport sweep.  We consider two variants of this approach: QPZ (QP Zero) and QPMP (QP Maximum Principle).
In QPZ, we constrain the coefficients of the solution to be nonnegative.  In QPMP, we further constrain the coefficients to satisfy an approximate discrete maximum principle.
In each spatial element, the objective function for the QP problem is the vector 2-norm of the difference between the coefficients of the local corrected and uncorrected angular fluxes.

In Section~\ref{sec:results}, we demonstrate the QPZ and QPMP methods on four 2-D problems, two of which are optically thick.  With the ZR, QPZ, or QPMP methods, negativities are eliminated when a positive (Bernstein) basis is used.  When a non-positive basis is used, negativities are mitigated but can still be present.  In terms of mitigating unphysical oscillations, the QPZ method performs slightly worse than the ZR method, while the QPMP method performs better than the ZR method.  In void regions, the QPMP scheme with the Bernstein basis produces a non-oscillatory solution that adheres to a discrete maximum principle.  In non-void regions, unphysical oscillations can still be present in the QPMP scheme, but they are still damped compared to the other methods.
In smooth problems (e.g., the smooth glancing problem in Section~\ref{sec:results}), the errors of the QP methods (as well as the ZR method) converge at the same rate as the uncorrected method when the mesh is sufficiently refined.

In problems with optically thick regions, acceleration is needed regardless of the fixup applied.  Diffusion synthetic acceleration (DSA) is a standard approach for overcoming the slow convergence of transport sweeps in this regime~\cite{larsen1984diffusion}.
However, fixups can alter the asymptotic limit of the transport system, rendering it inconsistent with the standard DSA system.
When fixups are present, application of standard DSA to transport sweeps near the thick diffusion limit results in a divergent iteration scheme.
On the other hand, our results show that the Variable Eddington Factor (VEF) method -- described in \cite{olivier2017variable,olivier2019high} as well as Section~\ref{sec:VEF} -- exhibits rapid convergence in both optically thick and thin problems when a fixup is applied.  For several of the problems considered in our results, VEF struggles to converge with uncorrected transport sweeps when negativities are present.
When a fixup (ZR, QPZ, or QPMP) is applied, VEF performs as expected, ensuring fast convergence in the thick diffusion limit.  The combination of VEF with the QPMP results in a method that (1) produces nonnegative solutions, (2) adheres to an approximate discrete maximum principle, and (3) converges quickly, regardless of the optical thickness of the domain.

\section{Spatial Discretization of the SN Transport Equation}

The monoenergetic, steady-state, isotropically scattering, discrete ordinates ($S_N$) linear Boltzmann transport equation is given by
\begin{gather}
	\bOmega_d \cdot \nabla_{\bx} \psi_d(\bx) + \sigma_t(\bx) \psi_d(\bx) = \frac{1}{4\pi} \sigma_s(\bx) \varphi(\bx) + q_d(\bx), \quad x \in \mathcal{D}, \label{eqn:transport} \\
	\psi_d(\bx) = \psi_{d,\text{inc}} (\bx), \quad \bx \in \partial \mathcal{D}, \quad \text{\bf{n}}(\bx)  \cdot \bOmega_d < 0 \,. \nonumber
\end{gather}
Here, \\
\begin{center}
\begin{tabular}{rl}
	$\mathcal{D} =$ & spatial domain with $\partial \mathcal{D}$ as its boundary, \\
	$\psi_d(\bx) =$ & angular flux in the direction of $\bOmega_d$ ($\bOmega_d \in \mathbb{S}^2$), \\
	$\sigma_s(\bx) =$ & macroscopic isotropic scattering cross section, \\
	$\sigma_t(\bx) =$ & total macroscopic cross section $\sigma_s(\bx) + \sigma_a(\bx)$, where \\
	$\sigma_a(\bx) =$ & macroscopic absorption cross section, \\
	$q_d(\bx) =$ & volumetric source in the direction of $\bOmega_d$, \\
	$\psi_{d,\text{inc}}(\bx) =$ & known, incident angular flux on the boundary $\partial \mathcal{D}$, \\
	$\text{\bf{n}}(\bx) =$ & outward-directed unit normal vector on $\partial \mathcal{D}$ at $\bx$,
\end{tabular} 
\end{center}
and the scalar flux $\varphi(\bx)$ is defined by the discrete ordinates integration of $\psi_d(\bx)$:
\begin{equation}
	\varphi(\bx) = \sum_{d=1}^{N_{\Omega}} \omega_d \psi_d(\bx). \label{eqn:scalarflux}
\end{equation}
In \refeqn{scalarflux}, the $\omega_d$ are angular quadrature weights defined such that $\omega_d > 0$ and
\begin{equation}
	\sum_{d=1}^{N_{\Omega}} \omega_d = 4 \pi .
\end{equation}
The notation here is consistent with that of~\cite{haut2019efficient,haut2018dsa}.

For an arbitrary-order DG spatial discretization, the weak form of \refeqn{transport} can be written as follows:
\begin{equation}
  \l( \bOmega_d \cdot \bG + F_d + M_t \r) \psi_d =
  M_s \varphi + q_d \,, \label{eqn:weakform}
\end{equation}
where $\bOmega_d \cdot \bG + F_d$ is the discrete, upwinded DG representation of the streaming term, $M_s$ and $M_t$ are block diagonal matrices (with each block corresponding to a spatial element) representing the scattering and total cross sections respectively,
$\psi$ and $\varphi$ are the solution vectors storing
the finite element coefficients,
and $q_d$ is a vector encapsulating both the spatial source $q(\bx)$ and the contribution from the incoming boundary flux $\psi_{d,\text{inc}}$.
In this paper, the variables $\psi(\bx)$ and $\varphi(\bx)$ are functions of the spatial variable $\bx$, while $\psi$ and $\varphi$ (without the $(\bx)$) refer to vectors of finite element coefficients.
We use $\bOmega_d \cdot \bG$ as a notational shorthand for the linear combination of matrices $\Omega_{d,x} G_x +\Omega_{d,y} G_y  + \Omega_{d,z} G_z $.
The matrix $\bG$ is block diagonal with respect to the spatial elements,
and $F_d$ captures all of the coupling between neighboring spatial elements.
The complete details of our DG discretization are omitted for brevity, but they can be found in~\cite{haut2019efficient,haut2018dsa}.  The finite element discretization in our research code is implemented using the infrastructure provided by the Modular Finite Element Library (MFEM)~\cite{mfem-library}.

\section{Source Iteration, Transport Sweeps, and Flux Fixups}

The standard procedure for solving the $S_N$ transport equations is to lag the scattering source and iterate:
\begin{equation}
  \l( \bOmega_d \cdot \bG + F_d + M_t \r) \psi_d^{(n+1)} =
  M_s \varphi^{(n)} + q_d \,. \label{eqn:SI}
\end{equation}
This process is known as source iteration~\cite{lewisbook}.
We solve \eqref{eqn:SI} by performing a ``transport sweep,'' which is effectively a block Gauss-Seidel iteration.
When no mesh cycles are present, performing a transport sweep for each angle $d$ inverts the operator on the left side of \refeqn{SI} exactly.
With curved meshes, however, the presence of mesh cycles breaks the lower block triangular structure of the operator.
As a result, transport sweeps only approximately invert the operator in \refeqn{SI} for curved meshes, and the number of iterations/sweeps required to converge is typically higher~\cite{haut2019efficient}.
Nonetheless, graph-based techniques can be used to efficiently determine ``quasi-optimal'' element orderings that mitigate the impact of the upper block triangular component on the convergence of transport sweeps~\cite{haut2019efficient}.
Although there are alternate means of solving~\refeqn{SI}, we focus on transport sweeps in this work.
Thus, the terms ``source iteration'' and ``transport sweep'' are used interchangeably throughout this paper.

In each transport sweep, local linear systems of the form
\begin{equation}
  A_{d,e} \psi_{d,e} = b_{d,e} \label{eqn:local_solve}
\end{equation}
are solved in each spatial element $e$ for each angle $d$.
Here, $\psi_{d,e}$ is a vector consisting of the DOFs of the angular flux in element $e$ for the $d$-th angle.
That is, $\psi_{d,e} = \left[ \psi_{d,e,1}, \ldots, \psi_{d,e,N_{\text{DOF}}} \right]^T$.
For angle $d$ and spatial element $e$,
\begin{equation}
  \tilde{\psi}_{d,e}(\bx) = \sum_{i=1}^{N_{\text{DOF}}} \psi_{d,e,i} u_{e,i}(\bx) \,,
\end{equation}
where $\tilde{\psi}_d(\bx)$ is the finite element approximation of $\psi_d(\bx)$, and
$\l\{ u_1(\bx), \ldots, u_{N_{\text{DOF}}}(\bx) \r\}$ are the finite element basis functions.
In physical space, the values of these basis functions are defined via a transformation from reference to physical space:
$u(\bx(\hat{\bx})) = \hat{u}(\hat{\bx})$.
The size of this local system is the number of DOFs in the local element from the DG discretization ($N_{\text{DOF}}$).
The matrix $A_{d,e}$ is the local block of the operator on the left side of \refeqn{SI}, corresponding to angle $d$ and element $e$.
$b_{d,e}$ is the local source obtained by summing the contribution from incoming angular fluxes (either from the boundary or other spatial elements), a lagged local scattering source, and the spatial source.

A local fixup scheme provides two additional advantages: it simplifies the preservation of local element balance (i.e., conservation of mass or the number of particles in each spatial element), and it ensures that the cost of applying the fixup is not significant.
Local balance is preserved in element $e$ for angle $d$ when the corrected solution $\bar{\psi}_{d,e}$ satisfies the following balance equation:
\begin{equation}
	\int_{\kappa_e} \l[ \bOmega_d \cdot \nabla_{\bx} + \sigma_t(\bx) \r] \bar{\psi}^{(n+1)}_d(\bx)  d\bx = \int_{\kappa_e} \l[ \frac{1}{4\pi} \sigma_s(\bx) \bar{\varphi}^{(n)}(\bx) + q_d(\bx) \r] d\bx \label{eqn:continuousbalance}
\end{equation}
Here, $\int_{\kappa_e} (\cdot) d\bx$ is a volume integral over spatial element $e$, and $\bar{\varphi}$ is defined in a manner analogous to that of $\varphi$ in \refeqn{scalarflux}:
\begin{equation}
  \bar{\varphi}^{(n)}(\bx) = \sum_{d=1}^{N_{\Omega}} \omega_d \bar{\psi}_d^{(n)}(\bx) \,.
\end{equation}

In this work, we only consider polynomial bases that sum to one:
\begin{equation}
  \sum_{i=1}^{N_{\text{DOF}}} u_i(\bx) = 1 \,. \label{eqn:unity}
\end{equation}
When such a basis is chosen, \refeqn{continuousbalance} can be expressed in discrete form as
\begin{equation}
  \label{eqn:balance}
  \mcone^T A_{d,e} \bar{\psi}_{d,e} = \mcone^T b_{d,e} \,.
\end{equation}
Here, $\mcone$ is a vector of 1's having the same size as $\bar{\psi}$.  Further comments on the choice of basis polynomials are provided in Section~\ref{sec:polynomials}, while a brief discussion on the sign of $\mcone^T b_{d,e}$ is provided in Section~\ref{sec:altfixup}.

When a fixup is local, sophisticated and balance-preserving schemes can be applied without significantly increasing the total computational cost.
Local fixups scale linearly with the number of spatial elements.
In contrast, global fixups are typically either simple (i.e., not requiring the solution of an optimization problem) or prohibitively expensive (optimization problems often have $O(n^3)$ or worse complexity).
In realistic problems, the number of spatial elements with negativities should be a small fraction of the overall problem, making the cost of a local fixup small compared to the cost of a transport sweep.

\subsection{``Zero and Rescale''}

In this paper, we will compare the results of our new methods to the simple, but efficient and relatively robust balance-preserving local fixup from  \cite{hamilton2009negative}.
As noted earlier, we refer to this approach as ``zero and rescale'' (ZR).
Following the notation of \refeqn{local_solve}, the fixup can be described as
\begin{subequations}
\begin{align}
	\bar{\psi}_{d,e,i}^{\text{Z}} &= \max \l\{ \psi_{d,e,i}, 0 \r\} \,, \label{eqn:ZR1} \\
	\bar{\psi}_{d,e,i}^{\text{ZR}} &= \cfrac{\mcone^T b_{d,e}}{  \mcone^T A_{d,e}  \bar{\psi}_{d,e}^{\text{Z}} } \bar{\psi}_{d,e,i}^{\text{Z}}  =  \cfrac{ \sum\limits_j b_{d,e,j} }{ \sum\limits_j \sum\limits_k \l( A_{d,e} \r)_{k,j} \bar{\psi}_{d,e,j}^{\text{Z}}   }\bar{\psi}_{d,e,i}^{\text{Z}}  \,. \label{eqn:ZR2}%
\end{align} \label{eqns:ZR}%
\end{subequations}%
Here, $i$ and $j$ index the local DOFs for each element-angle pair $d,e$.
As its name suggests, ZR zeros out all the negativities coefficients (\refeqn{ZR1}) and then rescales the resulting zeroed coefficients to ensure local element balance (\refeqn{ZR2}).
Local element balance is easily verified by replacing $\bar{\psi}_{d,e}$ in \refeqn{balance} with $\bar{\psi}_{d,e}^{\text{ZR}}$ defined in \refeqns{ZR}.
If the local solution in an element $\psi_{d,e}$ has no negativities, then $\bar{\psi}_{d,e}^\text{ZR} = \psi_{d,e}$ and no fixup is needed.

\subsection{Quadratic Programming Fixup}

We now describe a new fixup approach based on solving a local quadratic programming (QP) problem.
In this new approach, we replace $\psi_{d,e}$ with a corrected flux $\bar{\psi}_{d,e}$ that minimizes the quadratic objective function
\begin{equation}
	f(\bar{\psi}_{d,e}) = \l\| \bar{\psi}_{d,e} - \psi_{d,e} \r\|_2 \equiv \sum_{i=1}^{N_{\text{DOF}}} \l( \bar{\psi}_{d,e,i} - \psi_{d,e,i} \r)^2, \label{eqn:obj}
\end{equation}
subject to the constraints of particle balance (\refeqn{balance}) and
\begin{equation}
	\psi_{\text{max},d,e} \ge \bar{\psi}_{d,e,i} \ge \psi_{\text{min},d,e} \label{eqn:bounds}
\end{equation}
for all local DOFs $i$.
In words, we seek a flux-corrected local solution $\bar{\psi}_{d,e}$ that is ``as close as possible'' to the original uncorrected angular flux $\psi_{d,e}$, subject to local element balance and constraints on the finite element coefficients.

We consider two variants of the QP fixup which differ only in their definitions of $\psi_{\text{max}/\text{min},d,e}$.
The simpler variant QP Zero (QPZ) is similar to the ZR fixup -- it is designed to eliminate negativities:
\begin{subequations}
\begin{align}
	\psi_{\text{max},d,e}^\text{QPZ} &= \infty \,, \\*
	\psi_{\text{min},d,e}^\text{QPZ} &= 0 \,.
\end{align} \label{eqns:QPZ_bounds}%
\end{subequations}%
Both the QPZ and ZR fixup schemes eliminate negative solution coefficients and preserve local balance.
The primary difference between the two schemes is that QPZ also minimizes $\l\| \bar{\psi}_{d,e} - \psi_{d,e} \r\|_2$.

Our second QP-based fixup preserves an approximate discrete maximum principle (MP); we refer to this fixup as QPMP.
In the QPMP scheme, $\psi_{\text{max},d,e}$ and $\psi_{\text{min},d,e}$ are estimates of the maximum and minimum physically allowable angular flux in element $e$ for angle $d$.
To facilitate the derivation of these estimates, we first define the following scalar quantities:

\begin{center}
\begin{tabular}{rl}
	$\psi_{\text{in},\text{max},d,e} =$ & maximum incoming angular flux for direction $d$ from upwind spatial elements, \\
  $\psi_{\text{in},\text{min},d,e} =$ & minimum incoming angular flux for direction $d$ from upwind spatial elements, \\
   $b_{\text{max},d,e} =$ & the maximum source in direction $d$ in element $e$, \\
   $\sigma_{t,\text{max},e} =$ & the maximum value of the total cross section in element $e$, and \\
  $\Delta_{d,e} =$ & the maximum possible chord length in direction $d$ in element $e$.
\end{tabular}
\end{center}
We note that $b_{\text{max},d,e}$ includes both the spatial source $q_d$ and the lagged scattering source $M_s \varphi^{(n)}$.  For the latter three quantities, approximations can be used if it is not easy to calculate the true maximum values.  (Unless the element is a void, the constraints will only approximately enforce the discrete maximum principle, regardless of whether the latter three quantities are exact.)  In our numerical results, $\Delta_{d,e}$ is taken as the length of the diagonal of the bounding hypercube of the spatial element in physical space.  (In 1-D or 2-D problems, this distance must be divided by the cosine of the polar angle.)  Then, we can define the bounds for the QPMP method as follows: 
\begin{subequations}
	\begin{align}
		\psi_{\text{max},d,e}^\text{QPMP} &\equiv \psi_{\text{in},\text{max},d,e} + b_{\text{max},d,e} \Delta_{d,e} \,, \label{eqn:QPMPmax} \\*
		\psi_{\text{min},d,e}^\text{QPMP} &\equiv \psi_{\text{in},\text{min},d,e} e^{-\sigma_{t,\text{max},e} \Delta_{d,e}} \,.  \label{eqn:QPMPmin}
	\end{align} \label{eqns:QPMP_bounds}%
\end{subequations}%
Unlike \refeqns{QPZ_bounds}, \refeqns{QPMP_bounds} depend on the problem parameters (boundary conditions, total cross sections, and spatial sources) as well as the angular flux from neighboring upwind elements.
The physical intuition for \refeqns{QPMP_bounds} is as follows.
\refeqn{QPMPmax} adds $b_{\text{max},d,e} \Delta_{d,e}$ -- the maximum source multiplied by the maximum possible path length traversed through the element -- to the maximum incoming flux $ \psi_{\text{in},\text{max},d,e}$ to obtain an upper bound.
\refeqn{QPMPmin} decays the minimum incoming flux $\psi_{\text{in},\text{max},d,e}$ by $e^{-\sigma_{t,\text{max},e} \Delta_{d,e}}$, the factor by which a flux decays due to the collision cross section $\sigma_{t,\text{max},e}$ over a distance $\Delta_{d,e}$.
These bounds can be readily extended for multigroup problems in which sweeps are performed for one group at a time: one could simply add a group index to all the parameters in \refeqns{QPMP_bounds}.

The optimization problem described by Equations~\eqref{eqn:obj}, \eqref{eqn:balance}, and \eqref{eqn:bounds} is a special case of singly linearly-constrained, bounded quadratic programs (SLBQPs).
Solution methods for SLBQPs are described in detail in~\cite{dai2006new}.
The QPMP/QPZ methods have two convenient properties that make it easier to solve than the typical SLBQP: an objective function in the form of a squared Euclidean distance (\refeqn{obj}) and a linear constraint with positive weights (\refeqn{balance}).
For this particular type of SLBQP, the MFEM finite element library~\cite{mfem-library} has a special solver, adapted from~\cite{dai2006new} and based on code by Denis Ridzal at Sandia National Laboratories.
For well-posed problems (i.e., problems where it is possible for the local flux to satisfy both the linear balance constraint and the bounds), we find this solver to be robust and computationally efficient.

In future work, we may consider alternative objective functions.  For example, one could consider the $L_2$ norm or a 1-norm.  In the former case, the objective function would be more complex (a mass matrix would be introduced) and the current optimization solver would not be suitable.  The cost of solving this problem would be greater, but may not be prohibitive since it is still quadratic.  In case of a 1-norm, the optimization problem is no longer quadratic and, consequently, solving it may be significantly more costly.

\subsection{Choice of Basis Polynomials} \label{sec:polynomials}

In our finite element discretization, we consider two types of polynomial bases: the interpolatory Gauss-Lobatto (GL) basis, and the positive Bernstein basis~\cite{ainsworth2011bernstein}.  Neither of these bases are orthogonal, but both bases satisfy \refeqn{unity}.  With the GL basis, coefficients correspond to values of the solution at specific points in the element, i.e., the finite element nodes.
At each of the nodes, exactly one basis function attains a value of one,
while the other basis functions are zero.
However, nonnegative coefficients in the GL basis do not guarantee that
the solution is nonnegative everywhere.
When the polynomial degree is 2 or higher,
the solution can attain negative values at points that are between nodes,
even if the values at all the nodes are nonnegative.
Using the ZR fixup, or either of the QP fixups we have described,
will result in nonnegative solution values at the nodes,
but it does not guarantee nonnegativity at the intermediate points.

On the other hand, the Bernstein basis is not interpolatory: coefficients do not correspond to solution values.  \cite{anderson2017high} demonstrates several advantages of using Bernstein polynomials in applications where monotonicity is desired, while~\cite{Carnicer1993} describes their optimal ``shape-preserving'' properties.  One particularly useful property is that the Bernstein polynomials form a partition of unity: their values in the spatial element are bounded between 0 and 1, and they sum to 1 at every point in the spatial element (satisfying \refeqn{unity}).  Because of this, the local solution in a spatial element is always bounded above and below respectively by the maximum and minimum coefficient values of its Bernstein representation.  Thus, if the Bernstein coefficients of the solution satisfy the bounds in \refeqn{bounds}, the solution itself satisfies those bounds everywhere in the spatial element.
In other words, nonnegative coefficients guarantee nonnegative solutions, and applying the ZR and QPMP fixups to the Bernstein coefficients of $\psi_{d,e}(\bx)$ produces a solution that is nonnegative everywhere in the element.

However, the converse is not true: a nonnegative polynomial may have negative coefficients in the Bernstein basis.  As a result, there may be situations in which a fixup is applied even when the polynomial is positive, potentially leading to undesirable losses in accuracy.  We have not encountered this in practice, but there are several possible remedies that can be taken in the event that the loss of accuracy becomes an issue.  Instead of checking whether the Bernstein coefficients violate the prescribed bounds, one can iteratively compute the true minimum and maximum values of the local, uncorrected Bernstein polynomial representation and use these values to determine whether a fixup should be applied.  Alternatively, one could use a higher polynomial degree; it is shown in~\cite{diniz2016positive} that, for every nonnegative polynomial on a closed box, there exists a degree above which every Bernstein representation of it has strictly nonnegative coefficients.  Lastly, one could use a Gauss-Lobatto basis and introduce additional constraints to ensure positivity away from the interpolation points (e.g., enforce that the solution is positive on a grid of points selected finely enough that the derivatives of the polynomial can be used to bound its value away from 0).  We note that this would have the drawback of increasing the complexity of the quadratic programming problem, as it would no longer be singly linearly constrained.

\subsection{Fixups with Negative Sources} \label{sec:altfixup}

In a given element $e$, it is not possible to satisfy balance (\refeqn{balance}) with a nonnegative angular flux if $\mcone^T b_{d,e}$ is negative (i.e., if the integral of the source over the element $e$ is negative).
Although spatial sources should physically never be negative, acceleration steps (such as those described later in this paper) can often lead to negative scalar flux iterates and, consequently, negative scattering or emission sources.
Other non-spatial discretization errors such as approximations in angular expansions of scattering can also result in negative sources.\footnote{We thank reviewer 1 for pointing this out.}
In many of these cases, $b_{d,e}$ is only slightly negative and $\mcone^T b_{d,e} \ge 0$ still holds, meaning that it is possible to obtain a balance-preserving nonnegative angular flux.

However, in cases where $\mcone^T b_{d,e} < 0$, balance can no longer be satisfied on the element level with nonnegative angular fluxes, and the fixup options described in this paper (ZR, QPZ, or QPMP) will fail.
When this happens, alternative fixup schemes should be considered.
One potential approach would be to consider variants of the aforementioned fixups, applied on multiple elements at a time rather than individual elements.
Generally, even if $\mcone^T b_{d,e} < 0$, one can find a collection of nearby elements $\mathcal{N}_e$ (including $e$) such that
$$ \sum_{e' \in \mathcal{N}_e} \mcone^T b_{d,e'}  \ge 0 $$
Once a suitable $\mathcal{N}_e$ is found, one could sweep (without a fixup) on all the elements in $\mathcal{N}_e$ and then perform either the ZR, QPZ, or QPMP fixup on all of $\mathcal{N}_e$ at once, treating it as one large element.
In this approach, balance would be enforced on $\mathcal{N}_e$ rather than the individual elements.
There would be a single rescaling over all of $\mathcal{N}_e$ for the ZR fixup.
Likewise, for the QP fixups, there would be a single quadratic program with one balance constraint for all the elements of $\mathcal{N}_e$.
Although we have not tried this approach of fixing up multiple spatial elements yet, it may be an important part of our future studies.

\section{Variable Eddington Factor (Quasidiffusion) Method}  \label{sec:VEF}

In problems with optically thick regions, source iteration converges unacceptably slowly, and the $S_N$ transport operator becomes poorly conditioned.
A standard remedy for this poor conditioning is the Diffusion Synthetic Acceleration (DSA) method, which applies a diffusion-based correction to the scalar flux between source iterations~\cite{larsen1984diffusion}.
In~\cite{haut2018dsa}, a DSA preconditioner is derived for an $S_N$ transport operator with an arbitrary order DG discretization on a curved mesh.
It is shown in~\cite{haut2018dsa} that the resulting DSA-preconditioned discrete $S_N$ transport operator has a condition number independent of the mean free path.

However, the application of a fixup during a source iteration alters the transport operator in a nonlinear manner.  This resulting ``fixed-up'' transport operator is still highly ill-conditioned for optically thick problems, but DSA is no longer effective at accelerating the convergence of the solution of this new nonlinear operator.  Even if one linearizes this nonlinear transport operator at the (unknown) solution, the near-nullspace of the resulting linearized operator would differ from that of the original transport operator, and it would no longer be consistent with the DSA matrix derived for the original, uncorrected transport problem.
Moreover, because the nonlinear operator depends on the solution itself, such a linearization cannot readily be performed a priori.
In principle, one could devise an inner-outer iteration scheme in which the fixup/linearization is ``fixed'' in the inner loop so that the near-nullspace does not change from inner iteration to inner iteration.  In that case, a linear preconditioner can be devised for the inner iterations.  We may consider such an approach in our future work, as it is similar to the Newton-Krylov approach taken in~\cite{fichtl2010newton}.  However, an inner-outer approach would introduce significant computational cost, because there would be more transport sweeps and the preconditioner would have to be reformed at the beginning of each outer iteration. It is unclear whether such an approach would even converge.   

The Variable Eddington Factor (VEF) method~\cite{olivier2017variable,olivier2019high}, also known as quasidiffusion~\cite{mihalas,gol1964quasi}, provides an alternative, nonlinear remedy for the slow convergence of source iteration.
In VEF, the transport solution is coupled to a ``lower-order'' system\footnote{Here, ``lower-order'' refers to the size of the system, not the order of the spatial discretization scheme.} consisting of the first two angular moments of the $S_N$ transport equation (\refeqn{transport}).
The lower-order system is closed by an Eddington tensor (Eddington factor) defined using angular flux information from the most recent transport sweep.  Mathematically, the iteration scheme is described by the following equations:
\begin{subequations}%
\begin{gather}
  \bOmega \cdot \nabla_{\bx} \psi^{(l+\half)}(\bx) + \sigma_t \psi_d^{(l+\half)}(\bx) =
  \frac{1}{4\pi} \sigma_s \acute{\varphi}^{(l)}(\bx) + q_d(\bx) \,,
  \tag{\theequation} \label{eqn:VEF_transport} \\
  \nextParentEquation[eqns:VEF]
  \nabla_{\bx} \cdot \boldsymbol{J}^{(l+1)}(\bx) + \sigma_a \acute{\varphi}^{(l+1)}(\bx) =
  q_0(\bx) \,, \label{eqn:VEF1} \\
  \nabla_{\bx} \cdot \l( E^{(l+\half)}(\bx) \acute{\varphi}^{(l+1)}(\bx) \r) +
    \sigma_t \boldsymbol{J}^{(l+1)}(\bx) =
  \bq_1(\bx) \,. \label{eqn:VEF2}
\end{gather}%
\end{subequations}%
Here, $E(\bx)$ is the Eddington tensor, defined by
\begin{equation}
	E(\bx) \equiv \cfrac{\sum\limits_d \omega_d \bOmega_d \otimes \bOmega_d \psi_d(\bx) }{ \sum\limits_d \omega_d \psi_d(\bx) } = \frac{1}{\varphi} \sum\limits_d \omega_d \bOmega_d \otimes \bOmega_d \psi_d(\bx) \,, \label{eqn:Eddington}
\end{equation}
and $q_0$ and $\bq_1$ are the zeroth and first angular moments of the source $q_d$:
\begin{subequations}
	\begin{align}
		q_0(\bx) &= \sum_d \omega_d q_d(\bx) \,, \\
		\bq_1(\bx) &= \sum_d \omega_d \bOmega_d q_d(\bx) \,.
	\end{align} \label{eqns:q}%
\end{subequations}%
The details of VEF (e.g., the boundary conditions of the lower-order system) are omitted for brevity, but they can be found in~\cite{olivier2017variable,olivier2019high}.

In~\cite{olivier2017variable,olivier2019high}, it is shown that, like DSA-preconditioned source iteration, the VEF scheme converges rapidly for problems with optically thick regions (often in $O(10)$ iterations).  However, VEF differs from traditional acceleration methods such as DSA in two important aspects.  First, the use of the Eddington tensor for closure (\refeqn{Eddington}) makes the scheme nonlinear.  It is shown in~\cite{olivier2019high} that the Eddington tensor is significantly more converged than the scalar flux during source iterations.  VEF takes advantage of this by effectively ``transferring'' the convergence of the Eddington tensor to the scalar flux during each outer iteration.  Second, the VEF iteration scheme produces two solutions: one from the $S_N$ transport equations (\refeqn{VEF_transport}), and one from the moment equations (\refeqns{VEF}).
When the VEF and transport systems are independently discretized in space,
the scalar fluxes $\acute{\varphi}$ and $\varphi$ differ.
That is,
\begin{equation}
  \acute{\varphi} \ne \sum_d \omega_d \psi_d \,.
\end{equation}

In our work, the VEF system is also spatially discretized with finite elements.
The finite element approximation of $\acute{\varphi}$ resides
in the subspace $L^2(\mathcal{D})$, while the approximation of each
component of $\boldsymbol{J}$ resides in the subspace $H^1(\mathcal{D})$.
In order to maintain the order of accuracy of the transport system, the basis functions for the components of $\boldsymbol{J}$ must span the same polynomial space as the basis for the transport fluxes $\psi_d$.
However, the basis functions of the VEF scalar flux $\acute{\varphi}$ can span a polynomial space one degree lower than that used for the transport fluxes and still maintain the same order of accuracy~\cite{olivier2019high}.  In the results of this paper, the polynomial spaces of the VEF quantities are chosen as we have just described. 

For transport sweeps with fixup, the VEF scheme can be applied with a simple modification: we define the Eddington tensor $E$ using the fixed-up angular flux rather than the original angular flux.
Unlike the DSA system, the VEF system is nonlinear, and information regarding the angular flux is passed to the VEF system through the Eddington tensor.
In~\cite{olivier2017variable} and~\cite{olivier2019high}, we see that VEF can already tolerate differences between the transport and VEF systems on the order of the discretization error.  Because flux-fixups are also on the order discretization error, it should not be too surprising that the VEF method is able to robustly handle this additional inconsistency.

In the results, we see that, unlike DSA, the VEF method can be used in conjunction with flux-fixups.  The resulting combination is a method that (1) converges quickly despite the presence of optically thick regions, and (2) produces a transport angular flux that is everywhere positive and, in the case of QPMP, approximately maximum principle preserving.
In fact, VEF often struggles with negative angular fluxes.
When negative values are present in $\psi_d(\bx)$ and $\sum_d w_d \psi_d(\bx)$ approaches 0, the Eddington tensor is not well defined  can ``blow up'' numerically.
Whereas DSA is a linear preconditioner and can be applied to a source iteration scheme regardless of the sign of the angular fluxes, the VEF method often fails when negative angular fluxes are used to define the Eddington tensor.


\section{RESULTS} \label{sec:results}

\subsection{Glancing Void Problem}
\label{sec_glancing_jump}

The glancing void problem is a simple 2-D steady-state test problem on a unit square void with by a uniform 10$\times$10 spatial grid.
The incoming angular flux is 1 on the left edge for $\bOmega = \l( \frac{1}{\sqrt{3}}, \frac{1}{\sqrt{3}}, \frac{1}{\sqrt{3}}\r)$ and 0 for all other boundaries or angles.  The exact solution is simple, but not well-represented by polynomials:
\begin{equation}
	\psi_{\text{exact},d}(x,y) = \begin{cases}  1, & \text{if }y > x \text{ and } \bOmega_d = \l( \frac{1}{\sqrt{3}}, \frac{1}{\sqrt{3}}, \frac{1}{\sqrt{3}}\r) , \\
	0,  & \text{otherwise} \,. \end{cases}
\end{equation}
The DG basis functions in each element of this problem are \nth{4}-order (quartic) polynomials.  As noted in~\cite{maginot2017nonnegative}, the primary benefit of this test problem is that it is challenging for most discretization schemes.
Though the problem by itself is not realistic, voids do occur in practical problems and discontinuities are generally present due to misalignments in the direction of travel and the mesh edges (especially when the mesh is curved).
A similar problem is studied for the development of nonnegative methods in~\cite{maginot2017nonnegative}.

\begin{figure}[p]
\begin{center}
    \begin{subfigure}[t]{0.5\textwidth}
   \centering
        \includegraphics[scale=0.3]{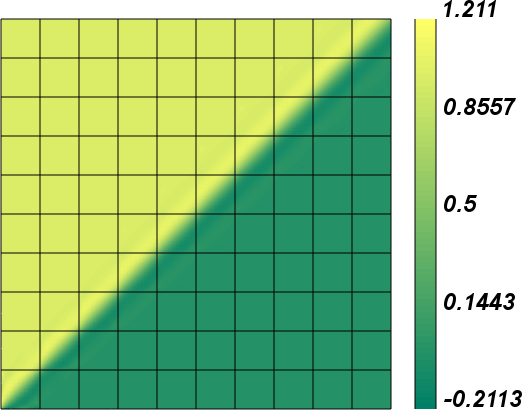}
        \caption{No fixup.}
    \end{subfigure} \vspace{0.5em} \\ \end{center}
    ~
    \begin{subfigure}[t]{0.5\textwidth}
   \centering
        \includegraphics[scale=0.3]{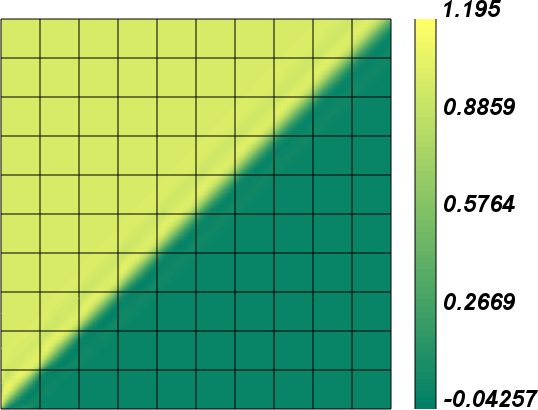}
        \caption{ZR, Gauss-Lobatto basis.}
    \end{subfigure}%
    ~
    \begin{subfigure}[t]{0.5\textwidth}
   \centering
        \includegraphics[scale=0.3]{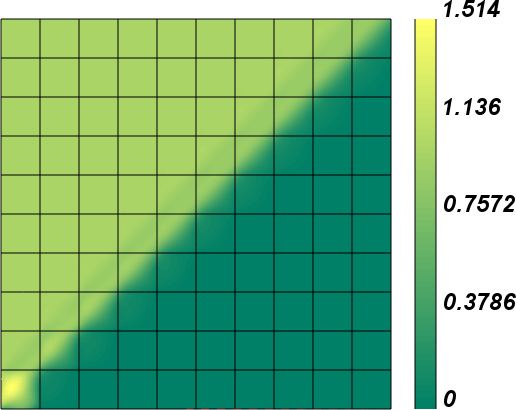}
        \caption{ZR, Bernstein basis.}
    \end{subfigure}\vspace{0.5em}  \\
    ~
    \begin{subfigure}[t]{0.5\textwidth}
        \centering
        \includegraphics[scale=0.3]{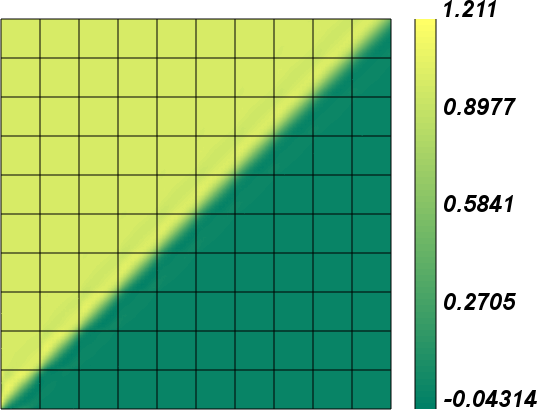}
        \caption{QPZ, Gauss-Lobatto basis.}
    \end{subfigure}%
    ~
    \begin{subfigure}[t]{0.5\textwidth}
        \centering
        \includegraphics[scale=0.3]{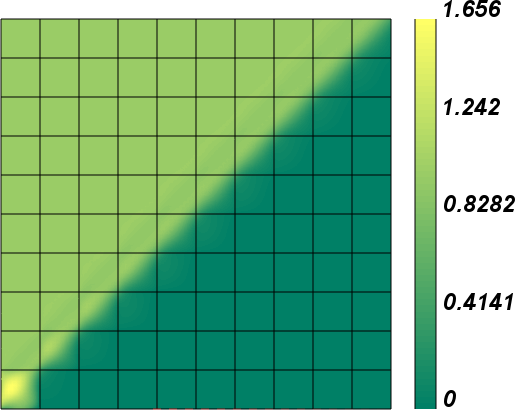}
        \caption{QPZ, Bernstein basis.}
    \end{subfigure}\vspace{0.5em} \\
    ~
    \begin{subfigure}[t]{0.5\textwidth}
        \centering
        \includegraphics[scale=0.3]{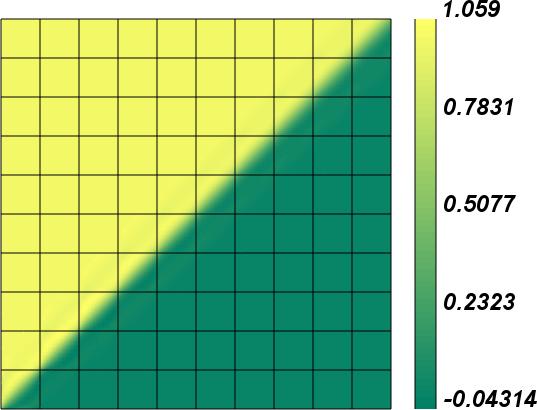}
        \caption{QPMP, Gauss-Lobatto basis.}
    \end{subfigure}
    ~
    \begin{subfigure}[t]{0.5\textwidth}
        \centering
        \includegraphics[scale=0.3]{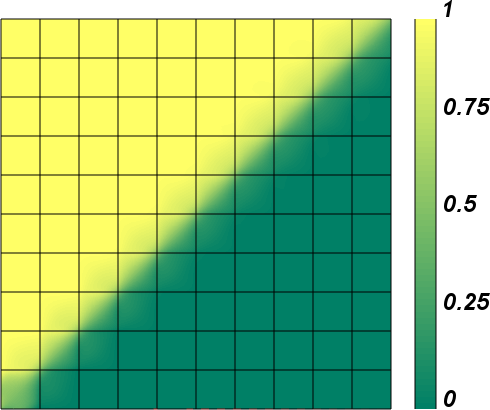}
        \caption{QPMP, Bernstein basis.}
    \end{subfigure}
    \caption{Plots of the scalar flux $\varphi$ in the glancing void problem on a straight mesh for various fixup options.} \label{fig:glancing_void_topdown}%
\end{figure}

\begin{figure}[p]
\begin{center}
    \begin{subfigure}[t]{0.5\textwidth}
   \centering
        \includegraphics[width=\linewidth]{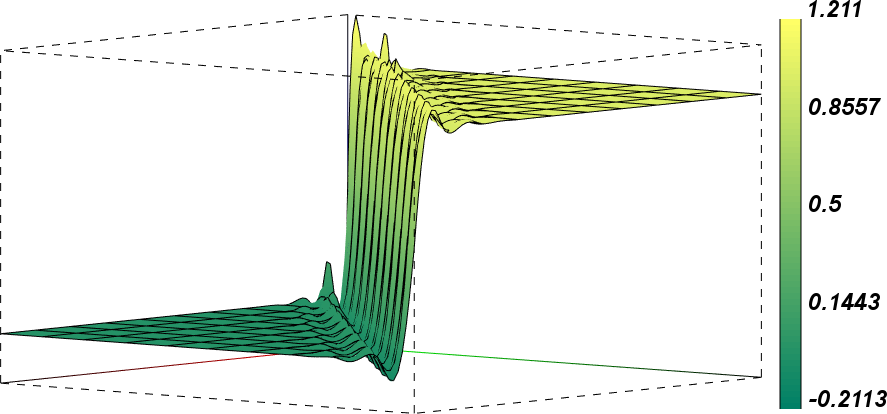}
        \caption{No fixup.}
    \end{subfigure} \vspace{0.5em} \\ \end{center}
    ~
    \begin{subfigure}[t]{0.5\textwidth}
   \centering
        \includegraphics[width=\linewidth]{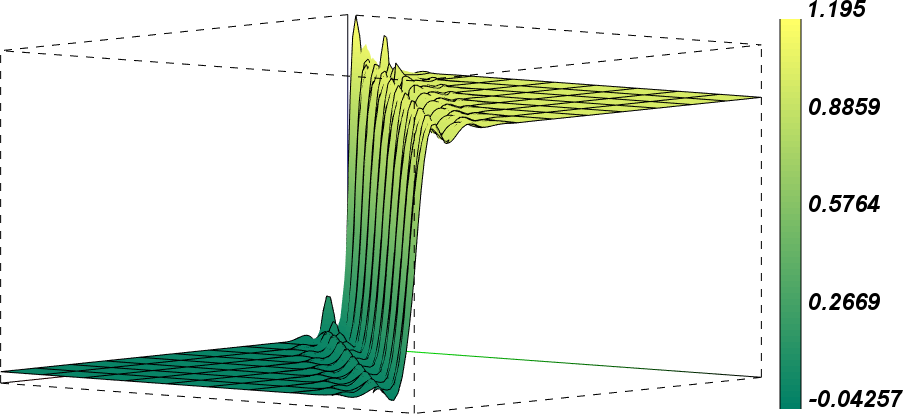}
        \caption{ZR, Gauss-Lobatto basis.}
    \end{subfigure}%
    ~
    \begin{subfigure}[t]{0.5\textwidth}
   \centering
        \includegraphics[width=\linewidth]{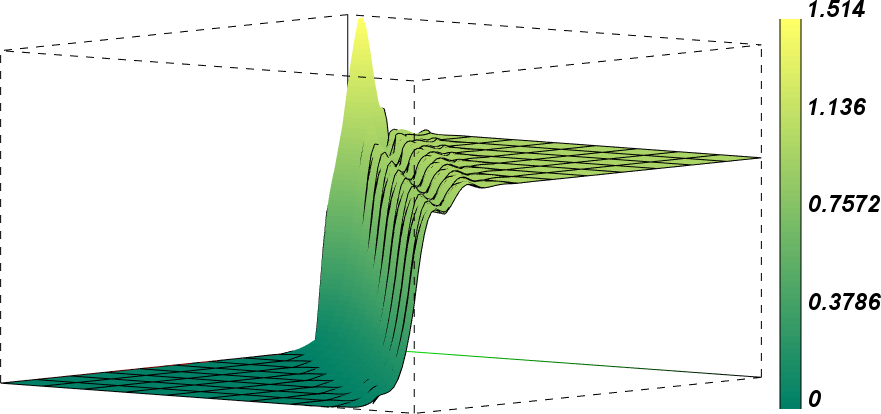}
        \caption{ZR, Bernstein basis.}
    \end{subfigure}\vspace{0.5em}  \\
    ~
    \begin{subfigure}[t]{0.5\textwidth}
        \centering
        \includegraphics[width=\linewidth]{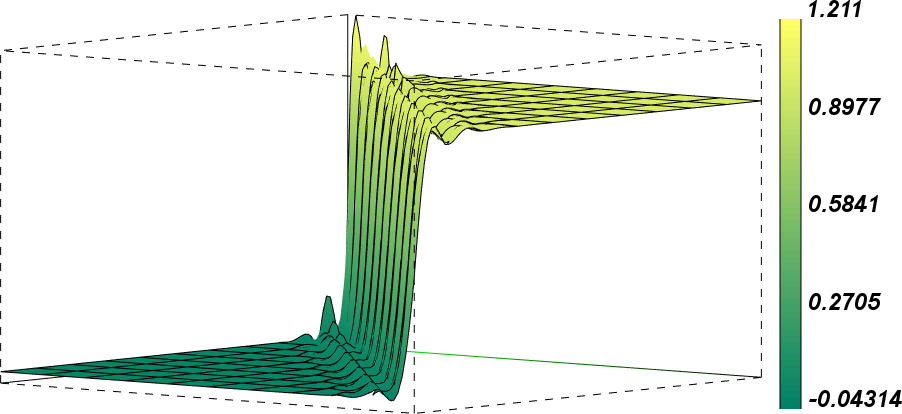}
        \caption{QPZ, Gauss-Lobatto basis.}
    \end{subfigure}%
    ~
    \begin{subfigure}[t]{0.5\textwidth}
        \centering
        \includegraphics[width=\linewidth]{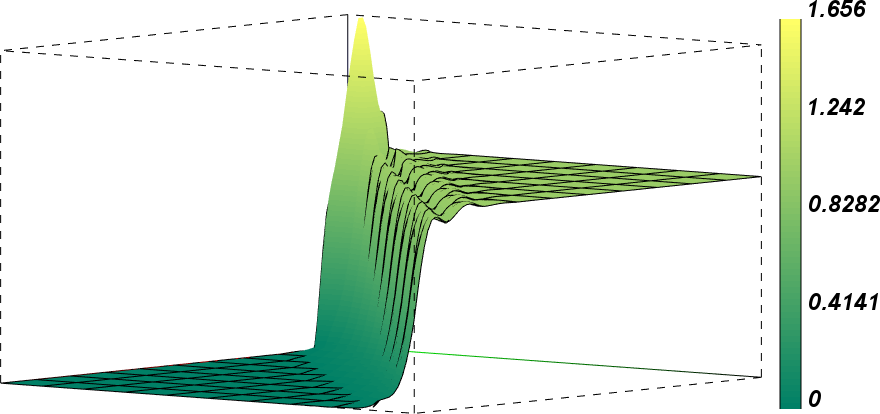}
        \caption{QPZ, Bernstein basis.}
    \end{subfigure}\vspace{0.5em} \\
    ~
    \begin{subfigure}[t]{0.5\textwidth}
        \centering
        \includegraphics[width=\linewidth]{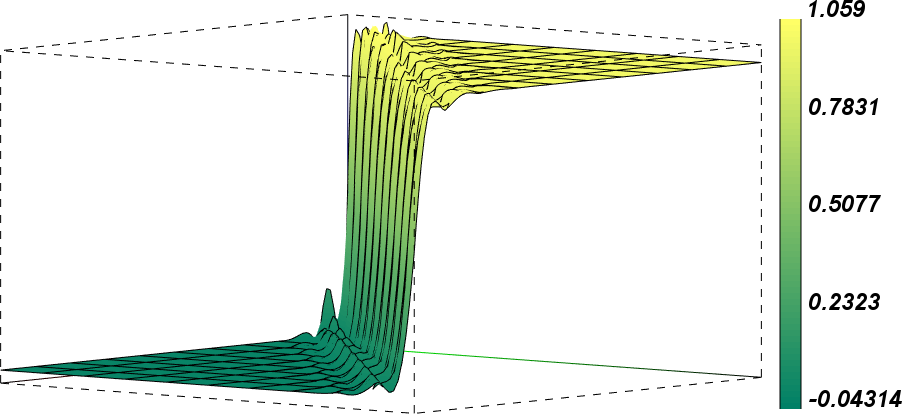}
        \caption{QPMP, Gauss-Lobatto basis.}
    \end{subfigure}
    ~
    \begin{subfigure}[t]{0.5\textwidth}
        \centering
        \includegraphics[width=\linewidth]{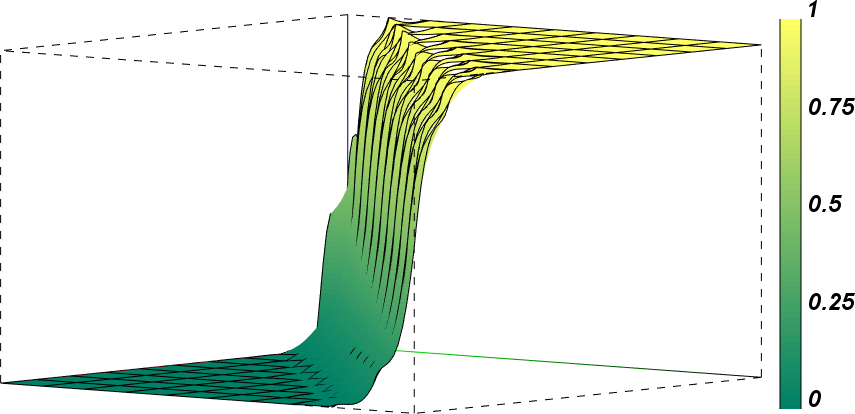}
        \caption{QPMP, Bernstein basis.}
    \end{subfigure}
    \caption{Perspective view of the plots in Figure~\ref{fig:glancing_void_topdown}, highlighting the oscillations along the solution discontinuity.} \label{fig:glancing_void}%
\end{figure}

Figures \ref{fig:glancing_void_topdown} and \ref{fig:glancing_void} show the results of the four fixup options for the glancing void problem, while Table~\ref{tab:error_glancing_void} lists the corresponding $L_2$ and $L_{\infty}$ errors.
In particular, Figure \ref{fig:glancing_void} provides a perspective view along the discontinuity, making it easier to see the prevalence and magnitude of the oscillations in each case.
We note that the uncorrected solutions are identical for the GL and Bernstein bases since the solution space spanned by the two sets of bases are the same.
This is not true for the corrected solutions, as both the zero and rescale and QP approaches alter the basis coefficients of the solution directly.

The QPZ method produces $L_2$ and $L_{\infty}$ errors that are similar to those of the ZR method, but it results in a larger unphysical peak near the origin.  For either choice of basis, the three fixup schemes produce similar $L_2$ errors.  When the fixup is applied to coefficients in the Gauss-Lobatto basis, the $L_2$ error is similar to that of the case without fixup.  When it is applied to coefficients in the Bernstein basis, the $L_2$ error is a bit higher as a result of the stricter bounds.

For either basis, QPMP mitigates the magnitude and presence of negativities, reduces the overshooting (how much the solution exceeds 1), and produces the most monotonic solution.  In the case of the Bernstein basis, the QPMP result lies strictly between 0 and 1, and it is monotonic in directions orthogonal to the discontinuity.  We note that this QPMP result with the Bernstein basis is the only fixed-up result that achieves the minimum possible $L_{\infty}$ error.  (Along the discontinuity, the analytical solution jumps from 0 to 1, and the error along this discontinuity must be 0.5 or greater since the finite element solution is a continuous polynomial within a spatial element.)  This lack of overshoots and undershoots is expected for the previously described reasons.  Unlike the Gauss-Lobatto basis, the Bernstein basis is positive, and a polynomial is bounded by its minimum and maximum Bernstein coefficients.  When we bound those coefficients using the QPMP or QPZ fixups, the resulting solution must fall strictly within those bounds.

\begin{table}[ht!]
  \centering
  \caption{\bf Summary of errors and overshoots/undershoots for the glancing void problem on a straight mesh}
  \begin{tabular}{rcccccccc}
  \toprule
  & &  \multicolumn{3}{c}{GL Basis} &&  \multicolumn{3}{c}{Bernstein Basis} \vspace{-2\aboverulesep}  \\ \cmidrule{3-5} \cmidrule{7-9}
   & No Fixup & ZR & QPZ & QPMP &&  ZR & QPZ & QPMP \vspace{-\aboverulesep} \\ \midrule
$L_{\infty}$ Error & 5.00e-1 & 5.50e-1 & 5.58e-1 & 5.08e-1 & & 1.25e+0 & 1.37e+0 & 5.00e-1\\
$L_2$ Error &  7.97e-2 & 7.81e-2 & 7.81e-2 & 8.10e-2 & & 1.18e-1 & 1.17e-1 & 1.13e-1 \\ \midrule
Max $\bar\psi$ & 1.21 & 1.20 & 1.21 & 1.06 & & 1.51 & 1.66 & 1 \\
Min $\bar\psi$ & -0.21 & -0.043 & -0.043 & -0.043 & & 0 & 0 & 0 \\ \bottomrule
  \end{tabular} \label{tab:error_glancing_void}
\end{table}

\FloatBarrier

\subsection{Glancing Void Problem with Curved Mesh}

\begin{figure}[p]
\begin{center}
    \begin{subfigure}[t]{0.5\textwidth}
   \centering
        \includegraphics[scale=0.3]{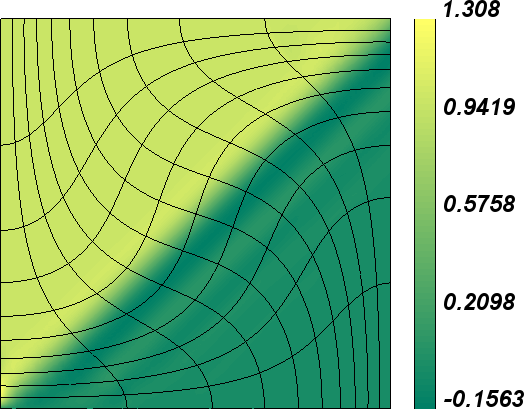}
        \caption{No fixup.}
    \end{subfigure} \vspace{0.5em} \\ \end{center}
    ~
    \begin{subfigure}[t]{0.5\textwidth}
   \centering
        \includegraphics[scale=0.3]{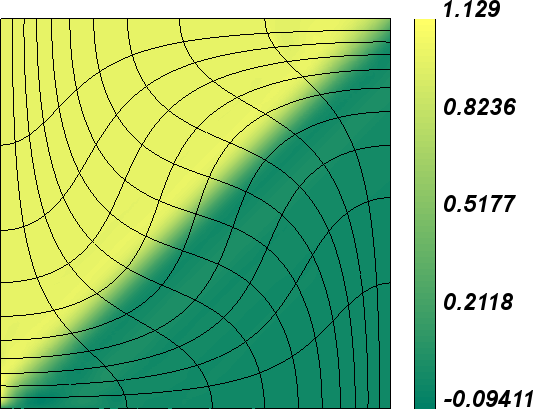}
        \caption{ZR, Gauss-Lobatto basis.}
    \end{subfigure}%
    ~
    \begin{subfigure}[t]{0.5\textwidth}
   \centering
        \includegraphics[scale=0.3]{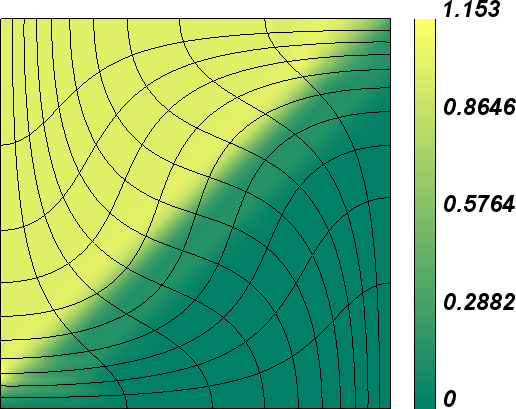}
        \caption{ZR, Bernstein basis.}
    \end{subfigure}\vspace{0.5em}  \\
    ~
    \begin{subfigure}[t]{0.5\textwidth}
        \centering
        \includegraphics[scale=0.3]{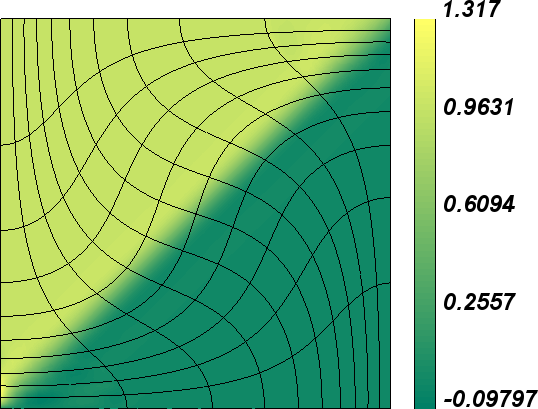}
        \caption{QPZ, Gauss-Lobatto basis.}
    \end{subfigure}%
    ~
    \begin{subfigure}[t]{0.5\textwidth}
        \centering
        \includegraphics[scale=0.3]{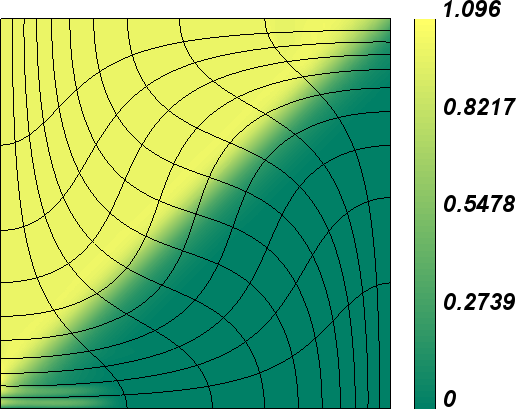}
        \caption{QPZ, Bernstein basis.}
    \end{subfigure}\vspace{0.5em} \\
    ~
    \begin{subfigure}[t]{0.5\textwidth}
        \centering
        \includegraphics[scale=0.3]{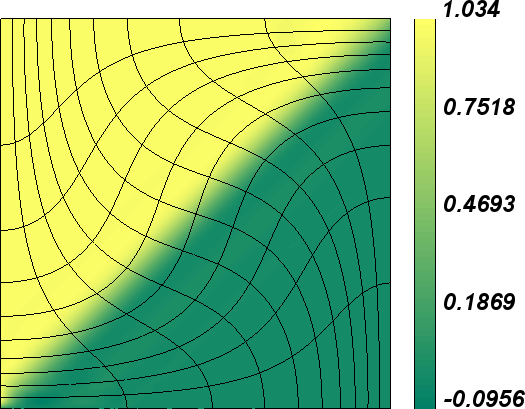}
        \caption{QPMP, Gauss-Lobatto basis.}
    \end{subfigure}
    ~
    \begin{subfigure}[t]{0.5\textwidth}
        \centering
        \includegraphics[scale=0.3]{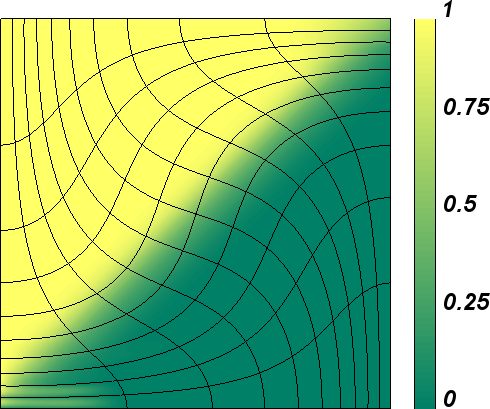}
        \caption{QPMP, Bernstein basis.}
    \end{subfigure}
    \caption{Plots of the scalar flux $\varphi$ in the curved mesh glancing void problem for various fixup options.} \label{fig:glancing_void_CM}%
\end{figure}

Next, we reconsider the glancing void problem, but with a curved mesh.
The mesh, which can be seen in Figure~\ref{fig:glancing_void_CM}, is generated by evolving the original 10$\times$10 uniform Cartesian mesh under the conditions of the 2-D Taylor-Green problem described in~\cite{dobrev2012high} to time $t = 0.4$ using the Lagrangian High-Order Solver (Laghos) hydrodynamics code~\cite{laghos}.
The purpose of this problem is to demonstrate that our fixup schemes work on curved meshes as well as straight meshes.

The $L_2$ and $L_{\infty}$ errors are tabulated in Table~\ref{tab:error_glancing_void_CM}, while top-view color-plots of the solutions are shown in Figure~\ref{fig:glancing_void_CM}.  From these results, we can see that there are some differences between the results of the straight mesh and curved mesh cases.  Because of the distortions in the mesh, none of the methods are able to achieve an $L_\infty$ error of 0.5.   Unlike the straight mesh case, there appear to be tangible benefits to using the QPZ fixup over the ZR fixup here, as the $L_2$ error, $L_{\infty}$ error, and the overshoot are all improved with the QPZ fixup compared to the ZR fixup.  We see some ``smearing'' or numerical diffusion in the elements near the bottom left corner of the domain (at the discontinuity of the incoming flux) due to the highly irregular and distorted nature of the elements there; this results in higher $L_2$ errors for all of the simulations.  Because of this smearing, the solution is not as monotonic as the straight mesh cases.

Nonetheless, the overall relative performance of the methods are similar qualitatively to that of the straight-mesh glancing void problem.
Undershoots and overshoots still occur when the fixups are applied to coefficients in the Gauss-Lobatto basis.
When the fixup is applied to coefficients in the Bernstein basis, negativities are eliminated in all of the fixup schemes.
However, as before, overshoots occur in the ZR and QPZ fixups regardless of basis, and only the QPMP fixup on a Bernstein basis has a solution that falls strictly between 0 and 1.

\begin{table}[ht!]
  \centering
  \caption{\bf Summary of errors and overshoots/undershoots for the glancing void problem on a curved mesh.}
  \begin{tabular}{rcccccccc}
  \toprule
  & &  \multicolumn{3}{c}{GL Basis} &&  \multicolumn{3}{c}{Bernstein Basis} \vspace{-2\aboverulesep}  \\ \cmidrule{3-5} \cmidrule{7-9}
   & No Fixup & ZR & QPZ & QPMP &&  ZR & QPZ & QPMP \vspace{-\aboverulesep} \\ \midrule
$L_{\infty}$ Error & 5.36e-1 & 5.47e-1 & 5.21e-1 & 5.26e-1 && 8.50e-1 & 6.49e-1 & 6.49e-1 \\
$L_2$ Error & 9.68e-2 & 1.00e-1 & 9.47e-2 & 1.03e-1 && 1.39e-1 & 1.23e-1 & 1.36e-1 \\ \midrule
Max $\bar\psi$ & 1.31 & 1.13 & 1.32 & 1.03 && 1.15 & 1.10 & 1 \\
Min $\bar\psi$ & -0.16 & -0.094 & -0.098 & -0.096 && 0 & 0 & 0\\ \bottomrule
  \end{tabular} \label{tab:error_glancing_void_CM}
\end{table}

\FloatBarrier

\subsection{Smooth Glancing Problem}

The smooth glancing problem is a modified version of a problem presented
in \cite{yuan2016high}.
Like the glancing void problem from Section \ref{sec_glancing_jump}, the domain
is a unit square, and there are no interior spatial sources.
The incoming flux is given by
\begin{equation}
	\psi_{\text{incoming},d}(x,y) = \begin{cases}  \sin^{12} (\pi y), & \text{if }x = 0 \text{ and } \bOmega_d = \l( 0.5, 0.1, \sqrt{0.74} \r) , \\
	0,  & \text{otherwise} \,.
	\end{cases}
\end{equation}
Here, we use \nth{3}-degree polynomials, and the domain is a pure absorber
with a cross section of 0.25.
This problem has a smooth analytic solution, but negativities and violations
of the maximum principle may occur in the numerical solution due to the
nature of the finite element approximation.
The purpose of this problem is to verify that our correction schemes preserve
high-order accuracy in problems with smooth solutions.
We note that the glancing void problem of the previous two sections is not suitable for this purpose since its analytic solution contains a discontinuity that is not mesh-aligned.
(Neither the uncorrected nor the corrected transport schemes can attain high-order accuracy in such a case.)

\begin{figure}[!ht]
	\begin{subfigure}{0.5\textwidth}
		\includegraphics[width=\textwidth]{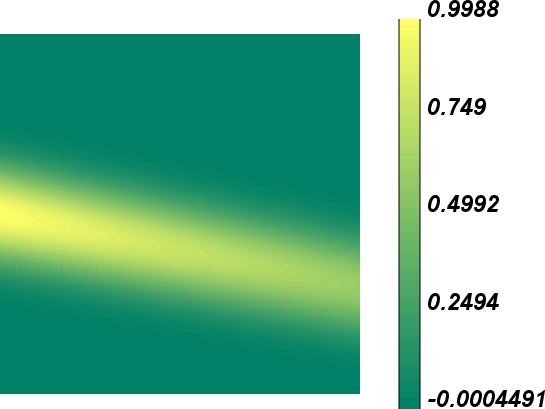}
		\caption{No fixup, $N_{x/y} = 10$.}
	\end{subfigure}%
	~
	\begin{subfigure}{0.5\textwidth}
		\includegraphics[width=\textwidth]{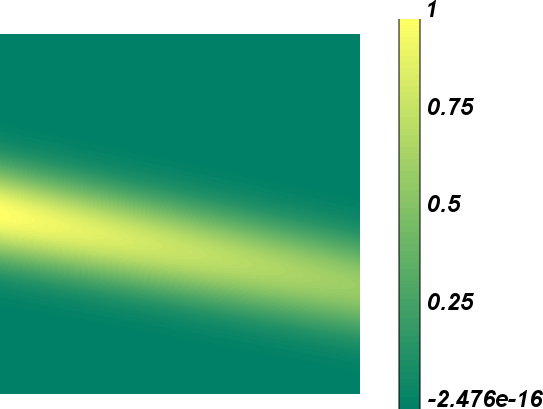}
		\caption{No fixup, $N_{x/y} = 160$.}
	\end{subfigure} \\
	\begin{subfigure}{0.5\textwidth}
		\includegraphics[width=0.95\textwidth]{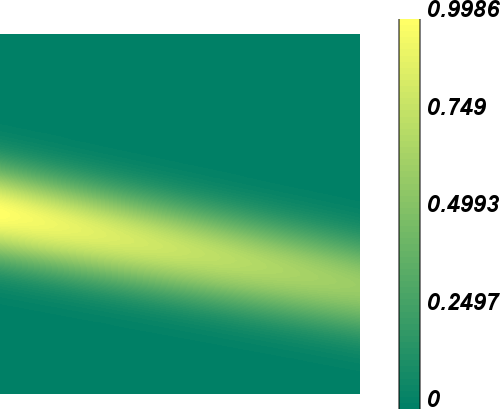}
		\caption{QPMP with Bernstein polynomials, $N_{x/y} = 10$.}
	\end{subfigure}%
	~
	\begin{subfigure}{0.5\textwidth}
		\includegraphics[width=\textwidth]{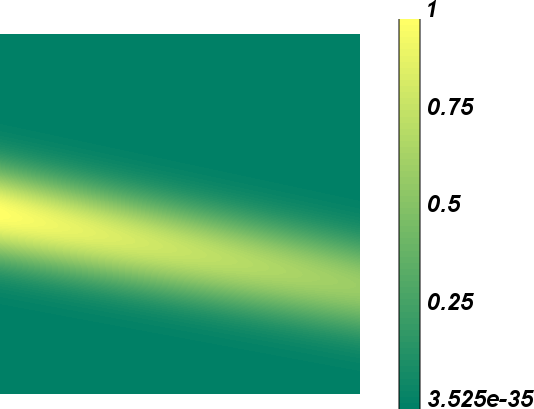}
		\caption{QPMP with Bernstein polynomials, $N_{x/y} = 160$.}
	\end{subfigure}
	\caption{Plots of the scalar flux $\varphi$ in the smooth glancing problem from~\cite{yuan2016high} for several selected cases.}
	\label{fig:smooth}
\end{figure}

Table~\ref{tab:order} provides a convergence study for the smooth glancing problem.  Several selected solutions are shown in Figure~\ref{fig:smooth}.  In this table, $N_{x/y}$ is the number of spatial elements per row or column, and the total number of spatial elements is $N_{x/y}^2$.  The grid is uniform in all cases.  The results here verify that the fixup methods have the same order of accuracy as the uncorrected method for sufficiently smooth problems.  This is because fixups are only performed when there are negativities or violations of the maximum principle, and the magnitudes of these violations are the same order as the overall error when the solution is sufficiently smooth.  We see that the ZR and QPZ methods stop making significant corrections after a few refinements of the grid (the errors are identical to the uncorrected case for $N_{x/y} \ge 40$), while the QPMP method continues to make small adjustments for larger $N_{x/y}$.

\begin{table}[ht!]
  \centering
  \caption{\bf $L_2$ errors for the smooth glancing problem}
  \begin{tabular}{ccccccccc}
  \toprule
  & &  \multicolumn{3}{c}{GL Basis} &&  \multicolumn{3}{c}{Bernstein Basis} \vspace{-2\aboverulesep}  \\ \cmidrule{3-5} \cmidrule{7-9}
  $N_{x/y}$  & No Fixup & ZR & QPZ & QPMP &&  ZR & QPZ & QPMP \vspace{-\aboverulesep} \\ \midrule
10 & 2.592e-04 & 2.652e-04 & 2.632e-04 & 3.275e-04 && 5.031e-04 & 3.202e-04 & 9.338e-04 \\
20 & 1.654e-05 & 1.654e-05 & 1.654e-05 & 2.773e-05 && 1.662e-05 & 1.656e-05 & 5.823e-05 \\
40 & 1.029e-06 & 1.029e-06 & 1.029e-06 & 1.987e-06 && 1.029e-06 & 1.029e-06 & 4.436e-06 \\
80 & 6.387e-08 & 6.387e-08 & 6.387e-08 & 1.153e-07 && 6.387e-08 & 6.387e-08 & 2.509e-07 \\
160 & 3.974e-09 & 3.974e-09 & 3.974e-09 & 7.322e-09 && 3.974e-09 & 3.974e-09 & 1.602e-08 \vspace{-\aboverulesep}  \\ \midrule
  \vspace{-2\aboverulesep} Order$^\dagger$ & 4.00 & 4.01 & 4.00 & 3.88 && 4.19 & 4.06 & 3.95 \\ \bottomrule
  \end{tabular} \label{tab:order} \\
${}^{\dagger}$The order of convergence is computed using a best-fit line through all of the data points.
\end{table}

\FloatBarrier
\subsection{Glancing Problem in the Thick Diffusion Limit}

Next, we revisit the glancing void problem from Section \ref{sec_glancing_jump}
and replace the void with a material that has the following cross sections:
\begin{subequations}
\begin{align}
	\sigma_t &= \frac{1}{\eps} \,, \\
	\sigma_a &= \eps \,, \\
	\sigma_s &= \frac{1}{\eps} - \eps \,.
\end{align}%
\end{subequations}%
This is the standard diffusion scaling, with $\epsilon \in (0,1]$.
Here, the thick diffusion limit corresponds to letting $\epsilon \to 0$.  Although the problem is highly scattering, the geometry, the boundary conditions, and the misalignments between the mesh and the directions of travel introduce discontinuities that lead to negative and oscillatory fluxes if no flux-fixup is applied.  On the other hand, as $\epsilon \to 0$, source iteration requires an arbitrarily large number of iterations to converge, and an acceleration scheme is needed.  The purpose of this problem is to study the compatibility of the ZR, QPZ, and QPMP fixups with DSA and VEF.  We use an \nth{8}-order Gauss-Legendre quadrature set for the $S_N$ angles and \nth{4}-degree Bernstein polynomials for the DG basis of the transport scalar flux.  The boundary conditions for this problem are given by:
\begin{equation}
	\psi_{\text{inc},d} = \begin{cases} 1, & x = 0, \Omega_{d,x} > 0, \Omega_{d,y} > 0, \\ 0, & \text{otherwise.} \end{cases}
\end{equation}

\begin{figure}[!ht]
	\begin{subfigure}{0.5\linewidth}
        \centering
        \includegraphics[width=\linewidth]{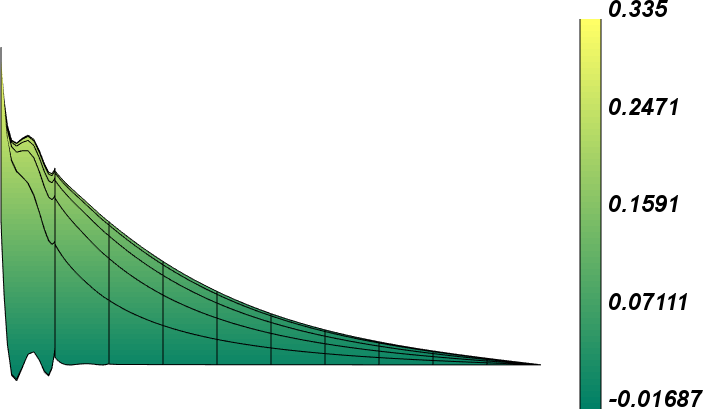}
        \caption{No fixup accelerated with DSA.} \label{fig:GS1}
	\end{subfigure}%
	~
	\begin{subfigure}{0.5\linewidth}
        \centering
        \includegraphics[width=\linewidth]{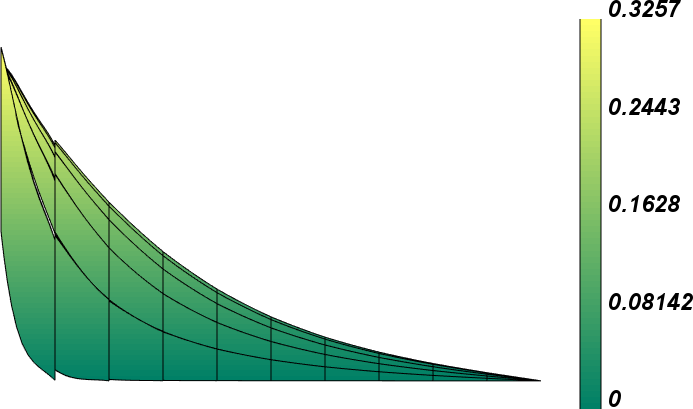}
        \caption{ZR accelerated with VEF.}  \label{fig:GS2}
	\end{subfigure}	\vspace{1em}\\
	\begin{subfigure}{0.5\linewidth}
        \centering
        \includegraphics[width=\linewidth]{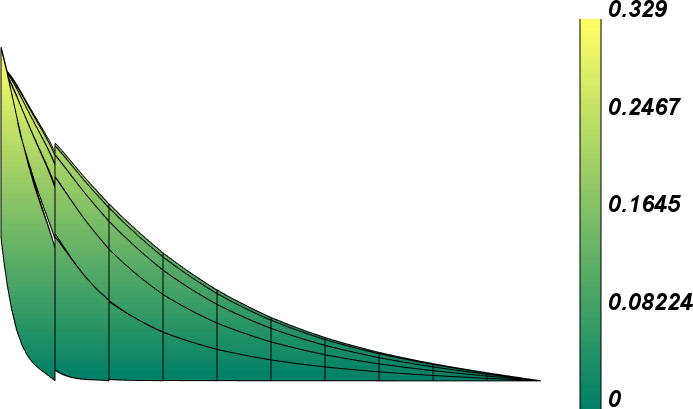}
        \caption{QPZ accelerated with VEF.}  \label{fig:GS3}
	\end{subfigure}%
	~
	\begin{subfigure}{0.5\linewidth}
        \centering
        \includegraphics[width=\linewidth]{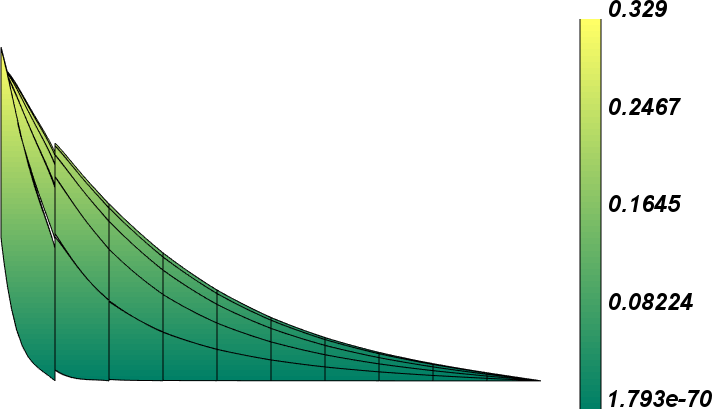}
        \caption{QPMP accelerated with VEF.}  \label{fig:GS4}
	\end{subfigure}%
\caption{Scalar flux ``side profiles'' for the $\epsilon = 10^{-3}$ cases that converged in Table~\ref{tab:GS_its}.  The horizontal axis is the $x$ dimension, the vertical axis is the scalar flux, and the axis perpendicular to the page is the $y$ dimension.  For the cases with VEF, the scalar flux shown here is obtained by integrating the transport angular flux over $4\pi$, not the scalar flux in the VEF system defined by \refeqns{VEF}.}
	\label{fig:GS}
\end{figure}

\begin{figure}[p]
	\begin{subfigure}{0.1\linewidth}
		~
	\end{subfigure}
	~
	\begin{subfigure}{0.45\linewidth}
		\centering
		No Fixup \hspace{0.2\linewidth}
	\end{subfigure}
	~
	\begin{subfigure}{0.45\linewidth}
		\centering
		QPMP \hspace{0.2\linewidth}
	\end{subfigure}\\
	~
	\begin{subfigure}{0.1\linewidth}
		$\epsilon = 0.5$		
	\end{subfigure}
	~
	\begin{subfigure}{0.45\linewidth}
		\centering
		\includegraphics[width=\linewidth]{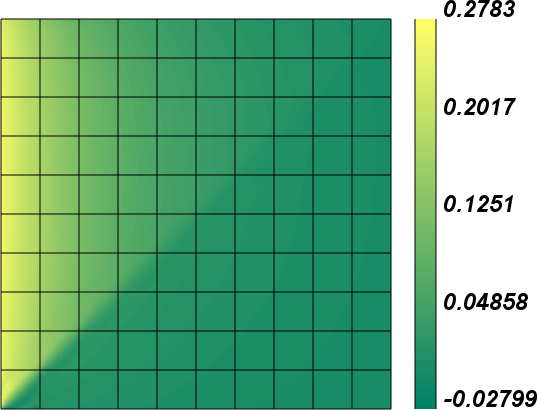}
	\end{subfigure}
	~
	\begin{subfigure}{0.45\linewidth}
		\centering
		\includegraphics[width=\linewidth]{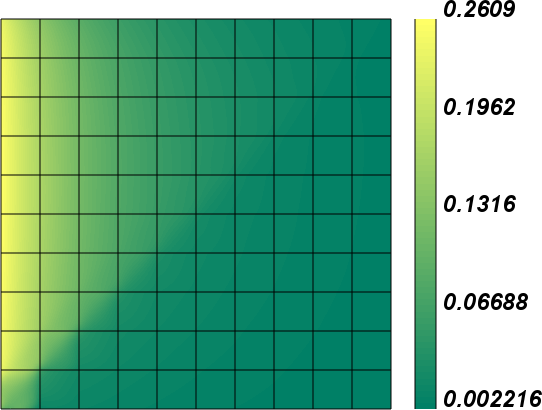}
	\end{subfigure} \vspace{0.5em}  \\
	~
	\begin{subfigure}{0.1\linewidth}
		$\epsilon = 0.01$		
	\end{subfigure}
	~
	\begin{subfigure}{0.45\linewidth}
		\centering
		\includegraphics[width=\linewidth]{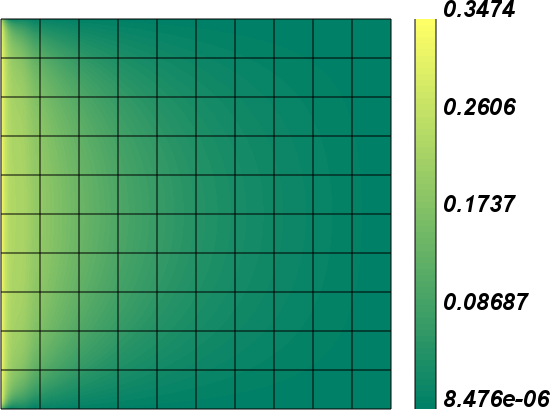}
	\end{subfigure}
	~
	\begin{subfigure}{0.45\linewidth}
		\centering
		\includegraphics[width=\linewidth]{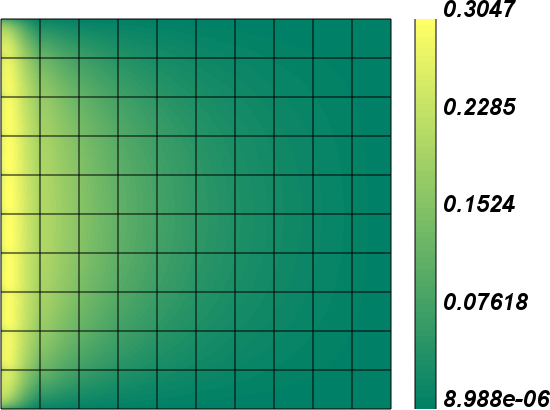}
	\end{subfigure} \vspace{0.5em}  \\
	~
	\begin{subfigure}{0.1\linewidth}
		$\epsilon = 0.0001$		
	\end{subfigure}
	~
	\begin{subfigure}{0.45\linewidth}
		\centering
		\includegraphics[width=\linewidth]{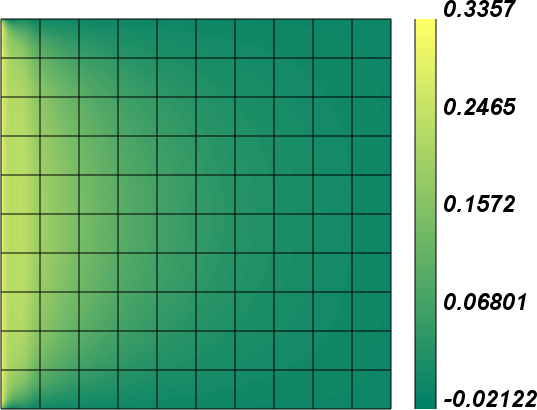}
	\end{subfigure}
	~
	\begin{subfigure}{0.45\linewidth}
		\centering
		\includegraphics[width=\linewidth]{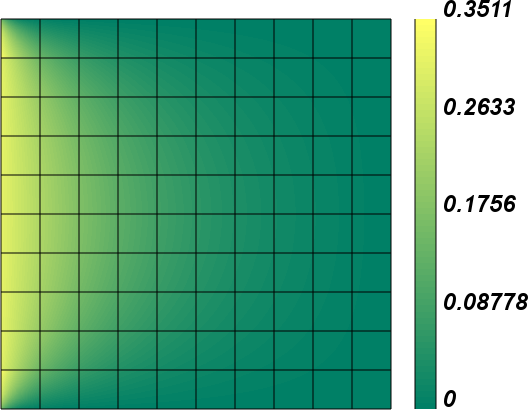}
	\end{subfigure}
	\caption{Plots of the scalar flux for the glancing problem for various values of $\epsilon$ with no fixup and with the QPMP fixup.  In the $\epsilon = 0.01$ case, the QPMP fixup alters the solution even though $\varphi$ is nonnegative.  This is because $\psi_d$ has negativities and discrete maximum principle violations for some angles $d$.}
	\label{fig:GS_eps}
\end{figure}

For $\epsilon = 10^{-3}$, Table~\ref{tab:GS_its} summarizes the performance
of each of the fixup options with each of the acceleration options.
Table~\ref{tab:GS_eps} presents a more detailed comparison between the
fixup-free and QPMP approaches for various values of $\epsilon$ for all acceleration options.
Solutions of select cases are shown in Figures \ref{fig:GS} and \ref{fig:GS_eps}.

\begin{table}[!ht]
	\caption{\bf Source iterations needed to converge the glancing problem with $\eps = 10^{-3}$ to a tolerance of $10^{-8}$.}
	\begin{center}%
	\begin{tabular}{ccc}%
	\toprule
	Fixup Scheme & Acceleration & Iterations Needed\\ \midrule	
	& None & ${}^\dagger2 \cdot 10^4$ \\
	None & DSA & 9\\
	& VEF & 72\\ \midrule
	& None  &  ${}^\dagger4 \cdot 10^4$\\
	ZR & DSA & Diverges \\
	& VEF & 7 \\ \midrule
	& None  & Stalls at $\sim10^{-3}$\\
	QPZ & DSA &  Diverges \\
	& VEF & 7 \\ \bottomrule
	& None  &  Stalls at $\sim10^{-3}$ \\
	QPMP & DSA &  Diverges \\
	& VEF & 7 \\ \bottomrule
	\end{tabular} \label{tab:GS_its} \\
	${}^\dagger$Estimated from the rate of convergence of first 10000 iterations.%
	\end{center}
\end{table}

\begin{table}[!ht]
	\centering
	\caption{\bf Source iterations needed to converge the glancing problem to a tolerance of $10^{-8}$ as $\eps \to 0$.}
	\begin{tabular}{c|ccc|ccc}
	\toprule
	Fixup Scheme & \multicolumn{3}{c|}{None} & \multicolumn{3}{c}{QPMP}\\
	Acceleration Scheme & None & DSA & VEF & None & DSA & VEF \\ \midrule
	$\eps = 0.5$ & 19 & 8 & 10 & 21 & 11 & 10 \\
	$\eps = 10^{-1}$ & 184 & 11 & 11 & 193 & 11 & 11 \\
	$\eps = 10^{-2}$ & 8531 & 18 &  Stalls at $\sim 10^{-1}$  & 8024 & Diverges & 11 \\
	$\eps = 10^{-3}$ & ${}^\dagger 2\cdot 10^4$ & 9 & 72 & Stalls at $\sim 10^{-3}$ & Diverges & 7  \\
	$\eps = 10^{-4}$ & ${}^\dagger 8\cdot {10^4}$ & 6 & 7 & Stalls at $\sim 10^{-3}$ & Diverges & 5  \\ \bottomrule
	\multicolumn{7}{l}{${}^\dagger$Estimated from the rate of convergence of the first 10000 iterations.}
	\end{tabular} \label{tab:GS_eps} 
\end{table}

There are several conclusions that can be drawn from these results.  First, it is clear that acceleration is needed by all schemes.  Without an acceleration scheme, the no-fixup case requires $\sim2 \cdot 10^4$ source iterations to converge for $\epsilon = 10^{-3}$.  With DSA, this number is reduced to 9.  The number of iterations required is similar for the ZR case.  For the QPZ and QPMP cases, convergence to a tolerance of $10^{-8}$ for $\epsilon = 10^{-3}$ does not appear to be possible; both cases stall at an error of $\sim 10^{-3}$.  This is because of insufficiently converged solutions to the quadratic program.  The maximum number of iterations in the quadratic program solver is set to 100 for all of the simulations.  After a certain number of source iterations, 100 iterations in the QP solver is not sufficient to generate a sufficiently converged solution to the quadratic program.  If we significantly increase the maximum number of iterations in the QP solver, it is possible for the QPZ and QPMP methods to converge without acceleration.  However, this would come at a great computational cost.  Thus, acceleration is needed for all of the fixup options when optically thick regions are present.

Furthermore, Table~\ref{tab:GS_its} shows that none of the fixup methods are compatible with DSA as $\epsilon \to 0$.
This is expected since the DSA scheme is derived to be consistent with the original transport equation, without fixup.  When a fixup is applied, the system no longer has the same asymptotic behavior, and it becomes inconsistent with standard DSA.  For $\epsilon \ge 10^{-2}$, it appears possible for a fixup scheme to converge with DSA.  However, all of the fixups schemes diverge with DSA for $\epsilon \le 10^{-3}$.  On the other hand, the fixup schemes converge with VEF for all values of $\epsilon$ we considered.
As $\epsilon \to 0$, we see in Table~\ref{tab:GS_eps} that the
QPMP method with VEF converges rapidly for all $\epsilon$.  Interestingly, the fixup-free case exhibits inconsistent behavior with VEF as $\epsilon \to 0$.  It diverges for the intermediate value $10^{-2}$, and it requires an unusually large number of iterations for $\eps = 10^{-3}$.
This unusual behavior likely stems from using an Eddington tensor derived from negative fluxes (see \refeqn{Eddington}); such an Eddington tensor can produce unphysical negative diffusion conductivity.

For this particular problem, the differences between the fixups do not appear to be significant.
Figure~\ref{fig:GS1} shows significant oscillations in the solution when no fixup is applied.
The reason for these oscillations is the same as the reason for the oscillations seen in the glancing void problems: a jump in the solution stemming from a discontinuity in the boundary condition.
In Figures \ref{fig:GS2}-\ref{fig:GS4}, we see that the oscillations are mostly removed when any of the three fixups are applied.
The peak values for the QPZ and QPMP are slightly closer to the peak value of the uncorrected solution, but otherwise the solutions appear qualitatively similar.

\begin{figure}[!ht]
	\begin{subfigure}{0.5\linewidth}
        \centering
        \includegraphics[width=\linewidth]{results_GS_QPMP_VEF_SNphi.png}
        \caption{Transport scalar flux $\varphi$}
	\end{subfigure}%
	~
	\begin{subfigure}{0.5\linewidth}
        \centering
        \includegraphics[width=\linewidth]{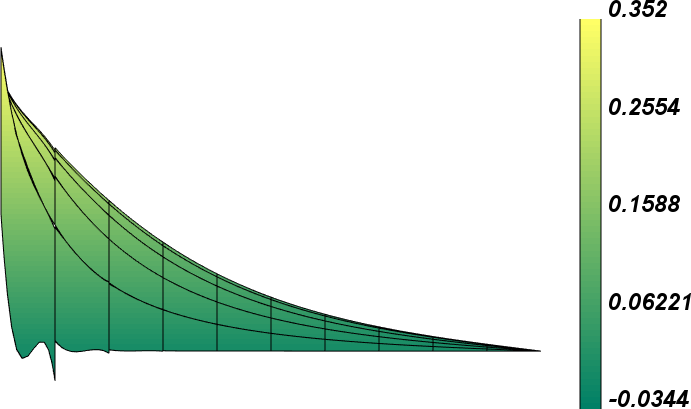}
        \caption{VEF scalar flux $\acute{\varphi}$}
	\end{subfigure}%
\caption{Comparison of the VEF and transport scalar fluxes for the QPMP fixup scheme with $\epsilon = 10^{-3}$.  As in Figure~\ref{fig:GS}, the horizontal axis is the $x$ dimension, the vertical axis is the scalar flux, and the axis perpendicular to the page is the $y$ dimension.}
	\label{fig:GS_compare}
\end{figure}

We recall that, in the case of the VEF-accelerated methods, there are two scalar fluxes to choose from: the transport scalar flux and the VEF scalar flux.  These two quantities differ on the order of the spatial discretization error, and deciding which of the two to use is still an open research question.  For the QPMP fixup with $\epsilon = 10^{-3}$, the difference between the VEF and transport scalar fluxes is shown in Figure~\ref{fig:GS_compare}.  Because no fixup is applied to the VEF system, there is no guarantee of positivity or discrete maximum principle preservation in the VEF scalar flux.  Indeed, we see that the VEF scalar flux in Figure~\ref{fig:GS_compare} has oscillations and negativities.  On the other hand, the transport scalar flux is positive since it follows directly from a discrete maximum principle preserving transport sweep.
As such, it appears that the transport scalar flux may be a more desirable option if a positive transport solution is needed in other equations in the application of interest (e.g., in the temperature equation in TRT).
In the problems of interest in this paper, the negativities from the VEF scalar flux $\tvp$ are not large enough to render the source negative in any of the local solves.  However, it is possible to derive a problem with spatial elements in which the spatial source and incoming angular flux are outweighed by the negativities in the scattering source from VEF.
This would break the iteration scheme, as all of the fixups described in this paper require a nonnegative total local source (i.e., the ZR, QPZ, and QPMP methods require $\mcone^T b \ge 0$ in \refeqn{local_solve}).
In such cases, one would either need to devise a fixup scheme for the VEF system, or consider an alternative fixup that can accommodate negative total local sources (see Section~\ref{sec:altfixup}). 
This issue will be further examined in our future work.

\FloatBarrier
\subsection{Tophat Pipe Flow Problem}

Finally, we consider the ``tophat'' or ``crooked pipe'' benchmark problem described in \cite{rss2015benchmark}.  A visualization of the geometry is shown below in Figure~\ref{fig:rss_geom}.  The problem consists of two materials: the wall and the pipe.  Their scattering cross sections are defined as follows:
\begin{subequations}
\begin{align}
	\sigma_{s,\text{pipe}} &= 0.2 \,, \\
	\sigma_{s,\text{wall}} &= 200 \,.
\end{align}%
\end{subequations}%
In either material, there is no ``real'' absorption.  However, we wish to emulate a backward Euler time step in our steady-state transport code.  To do so, we add a term of the form $\frac{1}{c \Delta t} \l[ \psi_d - \psi_{d,\text{old}} \r]$ to the transport equation.  We can account for this term in a steady-state context by incorporating it into the absorption and general source terms:
\begin{align}
	\sigma_a &= \frac{1}{c \Delta t} \,, \\
	q_d &= \frac{1}{c \Delta t} \psi_{d,\text{old}} \,.
\end{align}

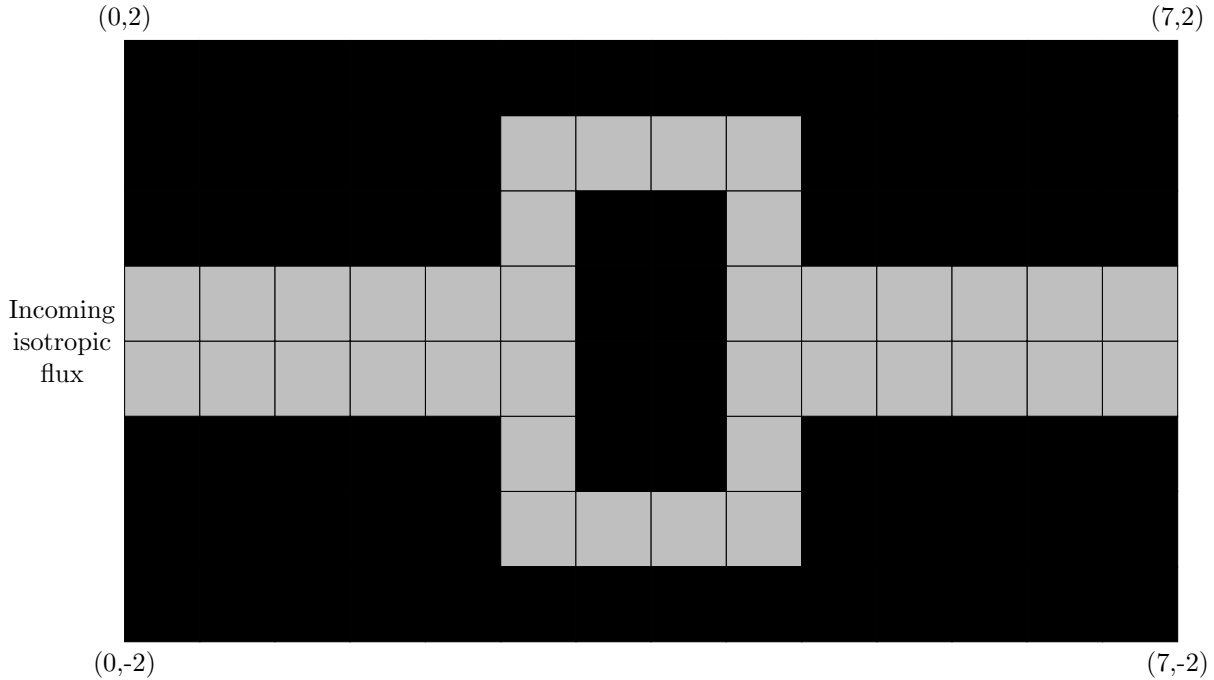
\begin{figure}[!ht]
	\centering
\begin{tikzpicture}
  \foreach \i in {\xMin,...,\xMax} {
  \foreach \j in {\yMin,...,\yMax} {
           \ifthenelse{\j > 2.9 \OR \j < -3.1 \OR \( \j > 0.9 \AND \i < 5. \)  \OR \( \j > 0.9 \AND \i > 8.9 \)
                         \OR \( \j < -1.1 \AND \i < 5. \)  \OR \( \j < -1.1 \AND \i > 8.9 \)
                         \OR \( \j < 2. \AND \j > -3.1 \AND \i < 8. \AND \i > 5.9 \) }
           {\filldraw [fill=black,draw=black] (\i,\j) rectangle ++(1.,1.);}
           {\filldraw [fill=lightgray,draw=black] (\i,\j) rectangle ++(1.,1.);}
  }
  }
  \node[left,align=center] at (0,0) {Incoming\\ isotropic\\ flux};
  \node[below,align=center] at (0,-4) {(0,-2)};
  \node[below,align=center] at (14,-4) {(7,-2)};
  \node[above,align=center] at (0,4) {(0,2)};
  \node[above,align=center] at (14,4) {(7,2)};
\end{tikzpicture}
	\caption{Tophat problem geometry.  The light gray region is the ``pipe,'' while the black region is the ``wall.''  This figure is a reproduction of Figure 1 from~\cite{rss2015benchmark}.} \label{fig:rss_geom} 
\end{figure}

To make the problem challenging, we consider a large time step $c \Delta t = 10^3$ with $\psi_{d,\text{old}} = 10^{-3}$ everywhere for all angles.  For this problem, we use an \nth{8}-order $S_N$ quadrature set and \nth{5}-degree Bernstein basis polynomials.  The spatial discretization is chosen to be relatively coarse in order to amplify the presence of negativities: the domain is divided into squares of width {0.5}.  There is an isotropic incoming flux of 1 on the left side of the pipe, and the boundary conditions are vacuum everywhere else:
\begin{equation}
	\psi_{\text{inc},d}(x,y) = \begin{cases}  1, & \text{if }x = 0 \text{ and } -0.5 < y < 0.5 , \\
	0,  & \text{otherwise} \,. \end{cases}
\end{equation}
Negativities and oscillations are present for high-order discretizations due to the material discontinuities and the sharp corners in the interface between the two materials.  This problem is particularly challenging for iteration schemes since there is both an optically thick material and an optically thin material.

\begin{figure}[!ht]
	\begin{subfigure}{0.5\linewidth}
        \centering
        \includegraphics[width=\linewidth]{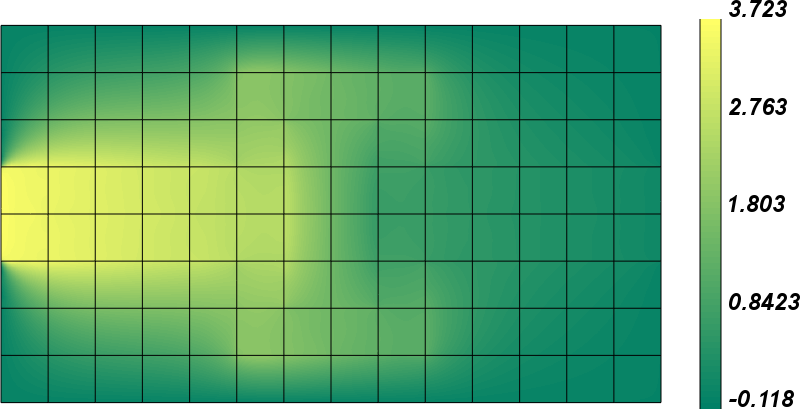}
        \caption{No fixup accelerated with DSA.}
	\end{subfigure}%
	~
	\begin{subfigure}{0.5\linewidth}
        \centering
        \includegraphics[width=\linewidth]{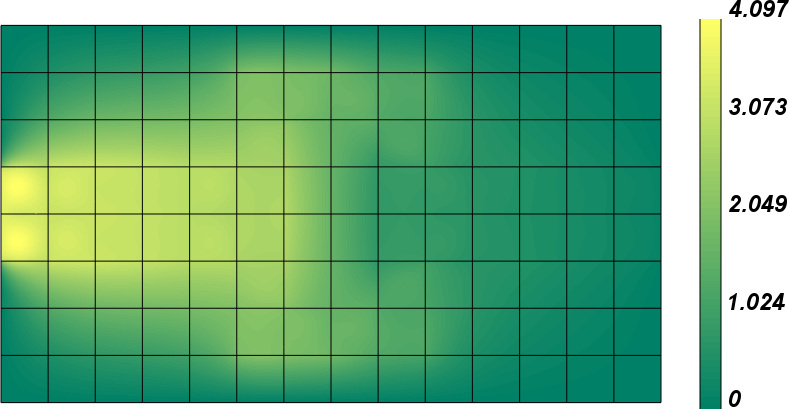}
        \caption{ZR accelerated with VEF.}
	\end{subfigure}	\vspace{1em}\\
	\begin{subfigure}{0.5\linewidth}
        \centering
        \includegraphics[width=\linewidth]{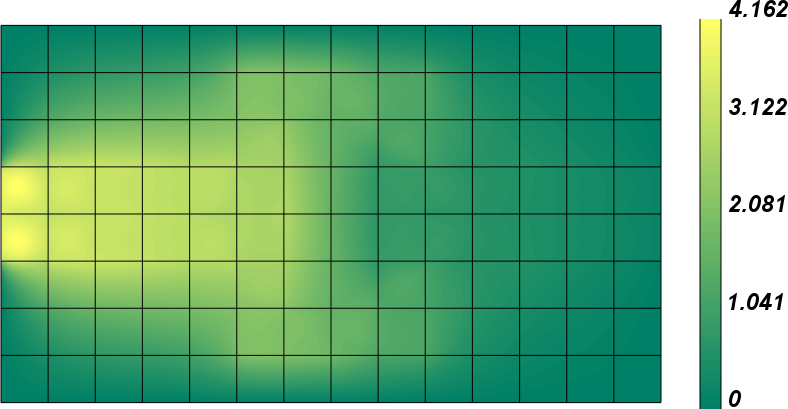}
        \caption{QPZ accelerated with VEF.}
	\end{subfigure}%
	~
	\begin{subfigure}{0.5\linewidth}
        \centering
        \includegraphics[width=\linewidth]{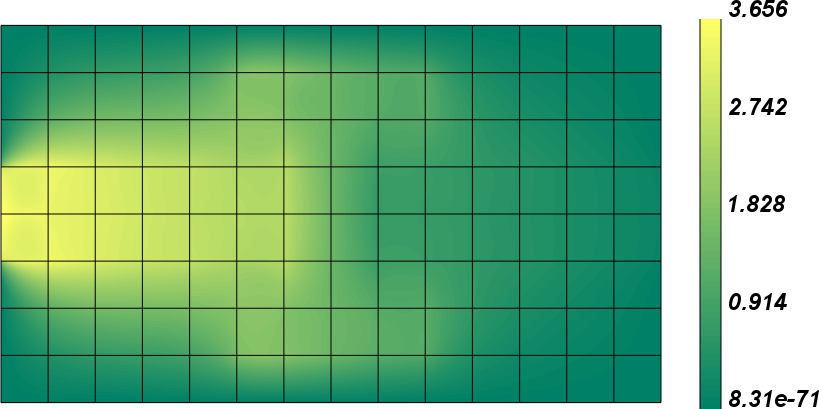}
        \caption{QPMP accelerated with VEF.}
	\end{subfigure}%
\caption{Plots of the scalar flux for the cases of the tophat pipe flow problem that converged in Table~\ref{tab:tophat}.  For the cases with VEF, the scalar flux shown here is the one obtained by integrating the transport angular flux over $4\pi$, not the scalar flux in the VEF system defined by \refeqns{VEF}.}
	\label{fig:tophat}
\end{figure}

As before, we consider the different combinations of fixup schemes and acceleration schemes.  The iteration counts and runtimes required are summarized in Table~\ref{tab:tophat}, and plots of the combinations that converged are shown in Figure~\ref{fig:tophat}.  The trends observed in these results are similar to those seen in the glancing problem in the diffusion limit.  Acceleration is needed to reduce the cost of solving the problem by orders of magnitude.  As before, we see that the convergence of the QP-based fixup schemes stalls without an acceleration due to insufficient convergence of the QP solver.  Because the fixed-up transport system does not have the same asymptotic diffusion limit as the fixup-free transport system, the iteration scheme diverges whenever DSA is used with any of the three fixup methods.  Moreover, without a fixup, the VEF-accelerated scheme does not converge.
Again, we believe this stems from defining an Eddington tensor with negative fluxes.

\begin{table}[!ht]
	\centering
	\caption{\bf Source iterations and runtimes needed to converge to a tolerance of $10^{-8}$ for the tophat problem.  The fixup fraction (fraction of elements requiring a fixup, averaged over all angles) is included in the right column.}
	\begin{tabular}{ccccc}
	\toprule
	Fixup Scheme & Acceleration & Iterations Needed & Runtime [s] & Fixup Fraction \\ \midrule	
	& None  & ${}^\dagger 4\cdot10^4$ & ${}^\dagger 4\cdot10^4$ & 0 \\
	None & DSA &  68 & 61 & 0\\
	& VEF & Stalls at $\sim 10^{-1}$ & -- & 0 \\ \midrule
	& None  & ${}^\dagger 3\cdot10^4$ & ${}^\dagger 3\cdot10^4$ & 0.44 \\
	ZR & DSA & Diverges & -- & --  \\
	& VEF & 63 & 119 & 0.25 \\ \midrule
	& None  & Stalls at $\sim 10^{-2}$ & -- & 0.82 \\
	QPZ & DSA & Diverges & -- & -- \\
	& VEF & 64 & 122 & 0.27 \\ \midrule
	& None  & Stalls at $\sim 10^{-2}$ & -- & 0.83  \\
	QPMP & DSA &  Diverges & -- & -- \\
	& VEF & 60 & 130 & 0.39 \\ \bottomrule
	\multicolumn{4}{l}{${}^\dagger$Estimated based on convergence progress in two hours of runtime.}
	\end{tabular} \label{tab:tophat}
\end{table}

As with the previous problem, VEF reduces the number of source iterations required for convergence by orders of magnitude for the cases where a fixup is used.
The number of iterations required for the successfully accelerated cases (DSA without fixup, VEF with fixup) are relatively high here compared to the iteration counts of the previous problem.
This is due to the large contrast in optical thickness between the two materials.
However, the iteration counts are similar to those seen in~\cite{rss2015benchmark}, and they are still significantly smaller than the iterations required for the unaccelerated cases.

Unlike the previous problem, there is a significant qualitative difference in the solutions between the different fixup options.  In Figure~\ref{fig:tophat}, we see that the ZR and QPZ methods result in unphysical peaks at the inlet of the pipe.
This is reminiscent of the unphysical peaks near the origin in the glancing void problem in Figure~\ref{fig:glancing_void}.
On the other hand, the solution from the QPMP fixup does not appear to have these unphysical peaks.  Its maximum value is much closer to the maximum value of the fixup-free solution, and the solution appears smoother near the pipe inlet.

\begin{figure}[!ht]
	\begin{subfigure}{0.5\linewidth}
        \centering
        \includegraphics[width=\linewidth]{results_tophat_QPMP_VEF_SNphi.png}
        \caption{Transport scalar flux.}
	\end{subfigure}%
	~
	\begin{subfigure}{0.5\linewidth}
        \centering
        \includegraphics[width=\linewidth]{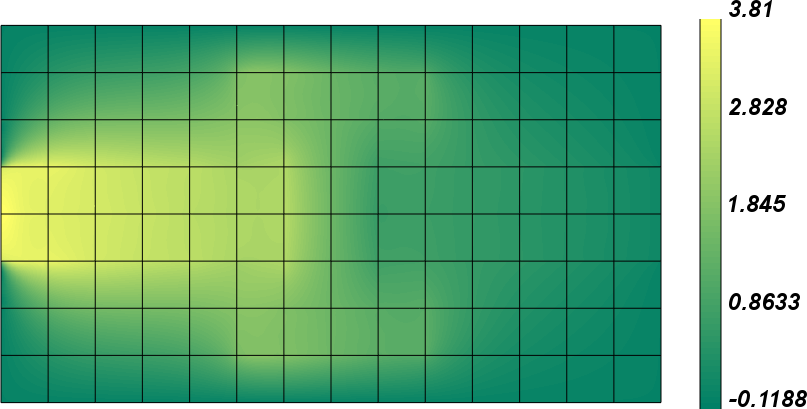}
        \caption{VEF scalar flux.}
	\end{subfigure}%
\caption{Comparison of the VEF and transport scalar fluxes for the QPMP fixup scheme for the tophat problem.}
	\label{fig:tophat_compare}
\end{figure}

Figure~\ref{fig:tophat_compare} shows the difference between the VEF and the
transport scalar fluxes for the QPMP fixup.
While the transport scalar flux is nonnegative by construction, the VEF scalar flux has significant negativities.  Although this does not occur in the problems considered in this paper, it is possible to construct a very difficult problem in which the right hand side of \refeqn{balance} (i.e., the total contribution of the scattering source and incoming fluxes from boundaries and other spatial elements) becomes negative due to large negativities in the VEF scalar flux $\acute{\varphi}$.  In such a case, a fixed-up transport sweep would not be possible since all of the fixup schemes described in this paper require the right hand side of \refeqn{balance} to be nonnegative.
(See Section~\ref{sec:altfixup} for a more thorough discussions on this.)
Though such cases may not be relevant in practice, we will consider the development of fixup schemes for the VEF scalar flux in the future in order to further improve the robustness of our iteration schemes.  For now, the fixed-up transport scalar flux will be used in multiphysics applications since it is nonnegative.

Finally, we consider the computational cost of applying the fixup.
In Table~\ref{tab:tophat}, we see that QPMP and QPZ with VEF only require
$\sim$9\% and $\sim$3\% more runtime, respectively, than the ZR method with VEF.
However, we do see a significant runtime increase when VEF is used instead of DSA.  This is because the VEF system is larger and more difficult to solve.  Currently, the direct solver SuperLU~\cite{li2005overview} is used for DSA and VEF.  It is possible to use other, more scalable solvers such as algebraic multigrid for SPD operators such as the DSA operator, but this is not possible for VEF since it is not SPD.  Improving the efficiency of VEF will be one of the primary objectives of our future work.

Because the efficiency differences between the QPMP and the no-fixup cases are
obscured by the performance differences between DSA and VEF, we took a second
approach to evaluate the computational burden of the fixup overhead.
With no acceleration, we ran transport sweeps with each fixup option for two hours on a single core to see how many source iterations each approach could complete in that time frame.  These results are presented in Table~\ref{tab:tophat_runtime}.
We see that, for the QPMP fixup, each source iteration requires about 31\% more runtime than the no-fixup case.
This is a significant increase in the computational cost, but it is certainly not computationally prohibitive.  Source iterations with the ZR fixup require only $\sim$5\% more computational time compared to the case with no fixup, while source iterations with QPZ require $\sim$9\% more.
We note that the problem we have considered is a particularly challenging one;
the spatial mesh is very coarse and corrections are needed in 83\% of the local
solves for the QPMP method (i.e., 83\% is the percentage of spatial elements requiring a fixup, averaged over all of the angles $d$).
In practice, this would indicate that the spatial mesh needs to be refined.
In more realistic problems, the number of elements requiring fixups should be
a much smaller fraction of the total count.
Thus, the results presented here should be viewed as a ``worst case'' scenario.
To reiterate, even when 83\% of the spatial elements require a fixup, the
computational overhead of the QPMP fixup is only a $\sim$31\% increase in the source iteration computational time.

\begin{table}[!ht]
	\centering
	\caption{\bf Number of unaccelerated source iterations completed in 2 hours of serial runtime.  The fixup fraction from Table~\ref{tab:tophat} is included here for the reader's convenience.}
	\begin{tabular}{cccc}
	\toprule
	Fixup Scheme & Source Iterations & Runtime per Source Iteration [s] & Fixup Fraction\\ \midrule	
	None & 6292 & 1.14 & 0 \\
	ZR & 6016 & 1.20 & 0.44 \\
	QPZ & 5828 & 1.24 & 0.82 \\
	QPMP & 4802 & 1.50 & 0.83\\ \bottomrule
	\end{tabular} \label{tab:tophat_runtime}
\end{table} 

\FloatBarrier
\section{Conclusions and Future Work}

We have developed a sweep-compatible fixup approach based on solving a local quadratic programming problem.
When combined with a positive basis such as the Bernstein basis, this approach eliminates negativities from the solution.
The bounds in the quadratic programming problem can be chosen either to prevent negativities (QPZ) or to go further and enforce adherence to an approximate discrete maximum principle (QPMP).  In void regions, the QPMP  method results in a strict adherence to the maximum principle.  In non-void regions, unphysical oscillations in the QPMP solution are dampened compared to either the uncorrected solution or the solution produced by the zero and rescale method described in~\cite{hamilton2009negative}.  Both methods preserve high-order accuracy in smooth problems because negativities and violations of the maximum principle are on the order of the discretization error.

As in the case where no fixup is applied, source iterations with any of the fixup schemes converge arbitrarily slowly in the optically thick diffusion limit.  Acceleration is needed for these cases.  However, the iteration schemes with fixup diverge when we apply consistent DSA schemes such as the one derived in~\cite{haut2018dsa}.  Fortunately, the VEF method does not have this issue.  By combining fixed-up transport sweeps with VEF, we can attain convergence rates which are independent of the mean free path.

The primary takeaway from this work is that the combination of the QPMP fixup with VEF results in a method that is nonnegative, adheres to an approximate discrete maximum principle, and converges rapidly regardless of the optical thickness of the domain.  There is a nontrivial increase in the computational cost of the source iterations when the QPMP method is used, but it is not prohibitive and it should be acceptable in most applications of interest.  Even in the tophat problem, where 83\% of the spatial elements needed a fixup, the computational overhead of performing a source iteration was only $\sim$31\%.  Future work includes examining the performance of our methods on a wider variety of practical problems, developing faster solvers for the VEF system, and extending the method to thermal radiative transfer (i.e., accounting for nonlinear iteration with the material energy balance equation).
\section*{ACKNOWLEDGMENTS}

Lawrence Livermore National Laboratory is operated by Lawrence Livermore National Security, LLC, for the U.S. Department of Energy, National Nuclear Security Administration under Contract DE-AC52-07NA27344.  This document (LLNL-JRNL-789678) was prepared as an account of work sponsored by an agency of the U.S. government.  Neither the U.S. government nor Lawrence Livermore National Security, LLC, nor any of their employees makes any warranty, expressed or implied, or assumes any legal liability or responsibility for the accuracy, completeness, or usefulness of any information, apparatus, product, or process disclosed, or represents that its use would not infringe privately owned rights.  Reference herein to any specific commercial product, process, or service by trade name, trademark, manufacturer, or otherwise does not necessarily constitute or imply its endorsement, recommendation, or favoring by the U.S. government or Lawrence Livermore National Security, LLC. The views and opinions of authors expressed herein do not necessarily state or reflect those of the U.S. government or Lawrence Livermore National Security, LLC, and shall not be used for advertising or product endorsement purposes.

 \setlength{\baselineskip}{12pt}
 \bibliographystyle{elsarticle-num}
 \bibliography{elsarticle}

\begin{thebibliography}{10}
\expandafter\ifx\csname url\endcsname\relax
  \def\url#1{\texttt{#1}}\fi
\expandafter\ifx\csname urlprefix\endcsname\relax\def\urlprefix{URL }\fi
\expandafter\ifx\csname href\endcsname\relax
  \def\href#1#2{#2} \def\path#1{#1}\fi

\bibitem{olivier2017variable}
S.~S. Olivier, J.~E. Morel, Variable eddington factor method for the {SN}
  equations with lumped discontinuous galerkin spatial discretization coupled
  to a drift-diffusion acceleration equation with mixed finite-element
  discretization, Journal of Computational and Theoretical Transport 46~(6-7)
  (2017) 480--496.

\bibitem{olivier2019high}
S.~S. Olivier, P.~G. Maginot, T.~S. Haut, High order mixed finite element
  discretization for the variable eddington factor equations, in: Proceedings
  of the International Conference on Mathematics and Computational Methods
  applied to Nuclear Science and Engineering (M\&C 2019), Portland, OR, 2019.

\bibitem{hamilton2009negative}
S.~Hamilton, M.~Benzi, J.~Warsa, Negative flux fixups in discontinuous finite
  element {$S_N$} transport, in: Proceedings of the International Conference on
  Mathematics, Computational Methods and Reactor Physics (M\&C 2009), Saratoga
  Springs, NY, 2009.

\bibitem{langer2015performance}
S.~Langer, I.~Karlin, V.~Dobrev, M.~Stowell, M.~Kumbera, Performance analysis
  and optimization for {BLAST}, a high order finite element hydro code, Tech.
  rep., Lawrence Livermore National Lab (LLNL), Livermore, CA (United States)
  (2015).

\bibitem{dobrev2012high}
V.~A. Dobrev, T.~V. Kolev, R.~N. Rieben, High-order curvilinear finite element
  methods for {L}agrangian hydrodynamics, SIAM Journal on Scientific Computing
  34~(5) (2012) B606--B641.

\bibitem{anderson2018high}
R.~W. Anderson, V.~A. Dobrev, T.~V. Kolev, R.~N. Rieben, V.~Z. Tomov,
  High-order multi-material {ALE} hydrodynamics, SIAM Journal on Scientific
  Computing 40~(1) (2018) B32--B58.

\bibitem{haut2019efficient}
T.~Haut, P.~Maginot, V.~Tomov, B.~Southworth, T.~Brunner, T.~Bailey, An
  efficient sweep-based solver for the {$S_N$} equations on high-order meshes,
  Nuclear Science and Engineering (2019) 1--14.

\bibitem{haut2018dsa}
T.~S. Haut, B.~Southworth, P.~G. Maginot, V.~Z. Tomov, {DSA} preconditioning
  for {DG} discretizations of {$S_N$} transport and high-order curved meshes,
  arXiv preprint arXiv:1810.11082 (2018).

\bibitem{wang2009high}
Y.~Wang, J.~C. Ragusa, A high-order discontinuous galerkin method for the {SN}
  transport equations on 2d unstructured triangular meshes, Annals of Nuclear
  Energy 36~(7) (2009) 931--939.

\bibitem{wang2010diffusion}
Y.~Wang, J.~C. Ragusa, Diffusion synthetic acceleration for high-order
  discontinuous finite element {SN} transport schemes and application to
  locally refined unstructured meshes, Nuclear Science and Engineering 166~(2)
  (2010) 145--166.

\bibitem{maginot2016high}
P.~G. Maginot, J.~C. Ragusa, J.~E. Morel, High-order solution methods for grey
  discrete ordinates thermal radiative transfer, Journal of Computational
  Physics 327 (2016) 719--746.

\bibitem{woods2018discrete}
D.~N. Woods, Discrete ordinates radiation transport using high-order finite
  element spatial discretizations on meshes with curved surfaces, Ph.D. thesis,
  Oregon State University (2018).

\bibitem{woods2019rz}
D.~N. Woods, T.~S. Palmer, {RZ} geometry discrete ordinates radiation transport
  using higher-order finite element spatial discretizations on meshes with
  curved surfaces, Journal of Computational and Theoretical Transport (2019)
  1--20.

\bibitem{lathrop1969spatial}
K.~Lathrop, Spatial differencing of the transport equation: positivity vs.
  accuracy, Journal of computational physics 4~(4) (1969) 475--498.

\bibitem{lewisbook}
E.~E. Lewis, W.~F. Miller, Computational Methods of Neutron Transport, American
  Nuclear Society, La Grange Park, IL, 1993.

\bibitem{wareing_exp}
T.~A. Wareing, An exponential discontinuous scheme for discrete-ordinate
  calculations in {C}artesian geometries, in: Proceedings of the Joint
  International Conference on Mathematical Methods and Supercomputing for
  Nuclear Applications, Saratoga Springs, NY, 1997.

\bibitem{maginot_LD}
P.~G. Maginot, J.~E. Morel, J.~C. Ragusa, A non-negative moment-preserving
  spatial discretization scheme for the linearized boltzmann transport equation
  in {1-D} and {2-D} {C}artesian geometries, Journal of Computational Physics
  231 (2012) 6801--6826.

\bibitem{maginot2017nonnegative}
P.~G. Maginot, J.~C. Ragusa, J.~E. Morel, Nonnegative methods for bilinear
  discontinuous differencing of the {SN} equations on quadrilaterals, Nuclear
  Science and Engineering 185~(1) (2017) 53--69.

\bibitem{maginot2015lumping}
P.~G. Maginot, J.~C. Ragusa, J.~E. Morel, Lumping techniques for {DEFM} {$S_N$}
  transport in slab geometry, Nuclear Science and Engineering 179~(2) (2015)
  148--163.

\bibitem{yuan2016high}
D.~Yuan, J.~Cheng, C.-W. Shu, High order positivity-preserving discontinuous
  {G}alerkin methods for radiative transfer equations, SIAM Journal on
  Scientific Computing 38~(5) (2016) A2987--A3019.

\bibitem{larsen1984diffusion}
E.~W. Larsen, Diffusion-synthetic acceleration methods for discrete-ordinates
  problems, Transport Theory and Statistical Physics 13~(1-2) (1984) 107--126.

\bibitem{mfem-library}
{MFEM}: Modular finite element methods library, \url{mfem.org}.
\newblock \href {https://doi.org/10.11578/dc.20171025.1248}
  {\path{doi:10.11578/dc.20171025.1248}}.

\bibitem{dai2006new}
Y.-H. Dai, R.~Fletcher, New algorithms for singly linearly constrained
  quadratic programs subject to lower and upper bounds, Mathematical
  Programming 106~(3) (2006) 403--421.

\bibitem{ainsworth2011bernstein}
M.~Ainsworth, G.~Andriamaro, O.~Davydov, {B}ernstein--{B}{\'e}zier finite
  elements of arbitrary order and optimal assembly procedures, SIAM Journal on
  Scientific Computing 33~(6) (2011) 3087--3109.

\bibitem{anderson2017high}
R.~Anderson, V.~Dobrev, T.~Kolev, D.~Kuzmin, M.~Q. de~Luna, R.~Rieben,
  V.~Tomov, High-order local maximum principle preserving ({MPP}) discontinuous
  galerkin finite element method for the transport equation, Journal of
  Computational Physics 334 (2017) 102--124.

\bibitem{Carnicer1993}
J.~M. Carnicer, J.~M. Pe{\~{n}}a,
  \href{https://doi.org/10.1007/BF02071384}{Shape preserving representations
  and optimality of the {Bernstein} basis}, Advances in Computational
  Mathematics 1~(2) (1993) 173--196.
\newblock \href {https://doi.org/10.1007/BF02071384}
  {\path{doi:10.1007/BF02071384}}.
\newline\urlprefix\url{https://doi.org/10.1007/BF02071384}

\bibitem{diniz2016positive}
M.~A. Diniz, L.~E. Salasar, R.~B. Stern, Positive polynomials on closed boxes,
  arXiv preprint arXiv:1610.01437 (2016).

\bibitem{fichtl2010newton}
E.~D. Fichtl, J.~S. Warsa, J.~D. Densmore, The {Newton-Krylov} method applied
  to negative-flux fixup in {SN} transport calculations, Nuclear Science and
  Engineering 165~(3) (2010) 331--341.

\bibitem{mihalas}
D.~Mihalas, Stellar Atmospheres, W. H. Freeman and Co, 1978.

\bibitem{gol1964quasi}
V.~Y. Gol'Din, A quasi-diffusion method of solving the kinetic equation, USSR
  Computational Mathematics and Mathematical Physics 4~(6) (1964) 136--149.

\bibitem{laghos}
\href{https://github.com/CEED/Laghos}{Laghos ({LAG}rangian {H}igh-{O}rder
  solver)}.
\newline\urlprefix\url{https://github.com/CEED/Laghos}

\bibitem{rss2015benchmark}
R.~P. Smedley-Stevenson, A.~W. Hagues, J.~Kópházi, A benchmark for assessing
  the effectiveness of diffusion synthetic acceleration schemes, in:
  Proceedings of the Joint International Conference on Mathematics and
  Computation, Supercomputing in Nuclear Applications and the Monte Carlo
  Method (M\&C 2015), Nashville, TN, 2015.

\bibitem{li2005overview}
X.~S. Li, An overview of {SuperLU}: Algorithms, implementation, and user
  interface, ACM Transactions on Mathematical Software (TOMS) 31~(3) (2005)
  302--325.

\end{thebibliography}

 \end{document}